\documentclass[sigconf,nonacm]{acmart}
\AtBeginDocument{%
  }

\usepackage{comment}
\usepackage{multirow}

\newif\ifDraft\Drafttrue
\ifDraft

\newcommand{\todo}[1]{
\textcolor{red}{TODO: #1\xspace}
}
\newcommand{\new}[1]{{\color{blue}{\bf #1}}}
\newcommand{\Red}[1]{{\color{red} #1}}
\newcommand{\tocite}[1]{\Red{CITE:\cite{#1}}} %missing cite
\newcommand{\toref}[1]{\Red{REF:\ref{#1}}} %missing ref
\newcommand{\XXX}[1]{\Red{\textbf{XXX[}#1\textbf{]}}} %XXX fix this

\else

\newcommand{\todo}[1]{}
\newcommand{\new}[1]{}
\newcommand{\tocite}[1]{}
\newcommand{\toref}[1]{}
\newcommand{\XXX}[1]{}

\fi

\usepackage{xcolor}

\definecolor{light-gray}{gray}{0.80}
\definecolor{light-blue-alt}{rgb}{0.648,0.817,0.872}
\definecolor{light-green}{rgb}{0.678, 1.0, 0.678}

\definecolor{light-blue}{HTML}{A5DAF5}

\newcommand{\niparagraph}[1]{\vspace{0.3em}\noindent\textbf{{#1}\hspace{0.5em}}}

\usepackage{enumitem}

\setlist[itemize]{leftmargin=6mm, topsep=5pt, noitemsep}
\newenvironment{CompactItemize}
  {\begin{itemize}[noitemsep,topsep=0pt,leftmargin=4mm]}
  {\end{itemize}}
\newcommand{\cmark}{\ding{51}}%
\newcommand{\xmark}{\ding{55}}

\usepackage{tikz}
\usepackage{amsmath}
\usepackage{graphicx}
\usepackage{xcolor}
\usepackage{booktabs}
\usepackage{pifont}
\usepackage{subcaption}
\usepackage{tabularx}
\usepackage{array}
\usepackage{makecell}
\usepackage{hyperref}

\begin{document}

\title[Leaky Language Models]{Leaky Language Models: Stealing Architecture and Inference Optimizations via Per-Token Timing}

\author{Sadegh Majidi}
\affiliation{%
  \institution{Purdue University}
  \city{West Lafayette}
  \state{IN}
  \country{USA}
}
\email{mmajidiy@purdue.edu}

\author{Niloofar Mireshghallah}
\affiliation{
  \institution{Carnegie Mellon University}
  \city{Pittsburgh}
  \state{PA}
  \country{USA}
}
\email{niloofar@cmu.edu}

\author{Kazem Taram}
\affiliation{
  \institution{Purdue University}
  \city{West Lafayette}
  \state{IN}
  \country{USA}
}
\email{kazem@purdue.edu}

\renewcommand{\shortauthors}{Majidi et al.}

\begin{abstract}
This work presents LeakyLMs, a set of attacks that 
leak proprietary model, architecture, and deployment information from production language models. LeakyLMs is the first to demonstrate that key model and deployment details can be inferred using only token generation timing, even when interacting through remote APIs.
LeakyLMs introduces two core attacks. The first attack targets inference optimizations and deployment strategies. For example, our attack detects whether a provider uses speculative decoding, a widely deployed inference-time optimization, and further identifies the context length of the draft model used in the pipeline.
Our measurements show that Google Gemini Flash 2.5 uses speculative decoding with a draft context window of approximately 128K tokens.
The second attack recovers key architectural properties, including the number of transformer layers, hidden dimension size, and number of attention heads. To achieve this, LeakyLMs builds a detailed and accurate model of token-generation timing on modern NVIDIA GPUs, characterizing how latency scales with model configuration and hardware parameters. The attack then performs a search over the architecture space using this timing model. In experiments with Llama models, the near-correct architectural configuration appears in the top-10 guesses more than 90\% of the time.
\end{abstract}


\maketitle

%-------------------------------------------------------------------------------
%%  Intro
%-------------------------------------------------------------------------------
%-------------------------------------------------------------------------------
\section{Introduction}

There is an ongoing global race across industry and governments to achieve superior AI performance and dominate the rapidly expanding AI market~\cite{tomshardware_huang2025, dw_china_ai_race}.
In this competitive landscape, model architecture and deployment strategies are valuable assets that provide competitive advantage. As a result, providers do not disclose the details of their most advanced models or their serving infrastructure, keeping them highly confidential~\cite{ntia_ai_system_disclosures_2024}.
For instance, although OpenAI recently released an open-source model (gpt-oss)~\cite{gpt-oss}, it scores significantly below their flagship model, GPT-5, on intelligence benchmarks~\cite{artificialanalysis_models}, and the architecture, training scale, and deployment details of GPT-5 remain largely undisclosed.

At the same time, providers heavily optimize the token-generation pipeline for low-latency streaming, with minimal tolerance for delay, to ensure high QoS and a seamless user experience~\cite{liu2024andesdefiningenhancingqualityofexperience}.
This exposes fine-grained, controlled, and accurate per-token generation timings of the model to external observers.
These per-token timings depend on the architectural parameters (e.g., depth, hidden dimension size, and attention heads), inference-time optimizations (e.g., speculative decoding), and deployment details (e.g., batching strategy, prompt caching, and hardware), many of which are proprietary and not publicly disclosed.

Therefore, careful observation of differential token timing, that is, comparing the latency between consecutive tokens, which cancels shared noise such as network delay,  allows an adversary to extract meaningful signals.
When combined with publicly known information, for example, that these systems are transformer-based or employ specific inference optimization techniques, this fine-grained timing data enables attackers to infer the otherwise undisclosed architectural and deployment details. 

This paper presents the first attacks of this kind. We demonstrate, for the first time, that it is possible to extract sensitive model and deployment information from production LLMs solely by observing token-generation timing.
In particular, we introduce two concrete attacks. 
The first targets deployment details, revealing whether a provider employs speculative decoding~\cite{leviathan2023spec}, a common inference-time optimization, and uncovering specific parameters of the underlying draft model used in this optimization.
Our second attack targets the internal model parameters such as the number of decoder layers, attention heads, and the hidden dimension size.

Just as hardware optimizations such as caching or speculative execution create timing channels, inference-time optimizations can introduce similar vulnerabilities~\cite{gu2025auditingpromptcachinglanguage}. 
Speculative decoding, for example, is an inference-time optimization that uses a smaller and faster draft model to produce multiple candidate tokens, and then invokes the main model to verify them. 
If the draft and main model disagree, the main model falls back to normal generation and produces the next token; if they agree, multiple tokens are accepted at once, improving throughput.

Our first attack leverages this behavior by crafting prompts that control how useful the draft model is. 
The draft model typically has a smaller context window, so when the correct prediction depends on distant context, the draft model will disagree with the main model. 
By varying prompt conditions to induce or avoid such disagreements, we produce measurable changes in token-generation timing that reveal the presence of speculative decoding, as well as the context length used by the draft model.

Our second attack targets model architectural parameters such as the number of decoder layers, hidden dimension size, and number of attention heads.
The attack is based on the intuition that token-generation timing forms a time series with predictable relationships to these architectural parameters. 
For instance, the hidden dimension size has a roughly quadratic impact on attention computation time, while the number of attention heads influences the latency linearly across tokens.

We first construct an accurate model of the timing behavior of transformer architectures to capture how architectural parameters affect per-token generation latency. 
We begin with an analytical examination of the transformer computation graph to derive the theoretical asymptotic relationships between runtime and key parameters. 
We then empirically refine this model by collecting runtime measurements across a range of configurations and fitting a linear regression model that combines the theoretical terms with empirically estimated constants, giving us a precise, explainable, and generalizable model of timing behavior.
We then use this model as an oracle to perform a grid search over possible architectural configurations, identifying the parameters that most closely reproduce the observed timing pattern of an unknown target model.

Our results show that both attacks successfully leak critical information through timing analysis. The first attack reveals that Google Gemini models Flash 1.5, Flash 2.5, and Flash 2.5 Lite all employ speculative decoding. Moreover, we are able to determine the context length of the draft models used in these systems---32 K, 128 K, and 128 K tokens, respectively. 

Our second attack infers key architectural parameters, including the number of layers, hidden dimension size, and number of attention heads of unseen models.
We show that this approach accurately models token-generation latency and enables recovery of architectural parameters across multiple transformer implementations, including baseline eager execution~\cite{att_all_you_nead, wolf-etal-2020-transformers}, FlashAttention2~\cite{dao2023flashattention2fasterattentionbetter}, and optimizations such as KV-cache~\cite{Jo2025FastKVKC,dai2019transformerxlattentivelanguagemodels}.
Our timing predictor generalizes to unseen architectures, achieving a normalized root mean squared error (NRMSE) of 0.12 for eager transformer implementations and an NRMSE of 0.188 for predicting prefill time of a model using FlashAttention2 with KV caching. 
Using this predictor, we recover the number of layers of unseen architectures within $\pm 1$ of the true value, with top-5 accuracy of 86.15\%, 97.69\%, and 83.78\% for eager, FlashAttention2, and FlashAttention2 with KV caching, respectively.
Finally, we demonstrate that the attack can recover parameters of an unknown model hosted on an inference serving platform (\textit{Weights \& Biases}). The predicted hidden dimension matches the ground truth, and the correct number of layers appears as the fourth-ranked candidate in our classifier.

\niparagraph{Responsible Disclosure.} We disclosed our speculative decoding attack to Google on Nov 7, 2025. Google acknowledged our findings on Jan~29,~2026.

%-------------------------------------------------------------------------------
%%  Background
%-------------------------------------------------------------------------------
%-------------------------------------------------------------------------------
\section{Background}
\label{sec:background}
%-------------------------------------------------------------------------------

\subsection{Transformer Architecture}

\begin{figure}[t]
    \centering
    \includegraphics[width=1\linewidth]{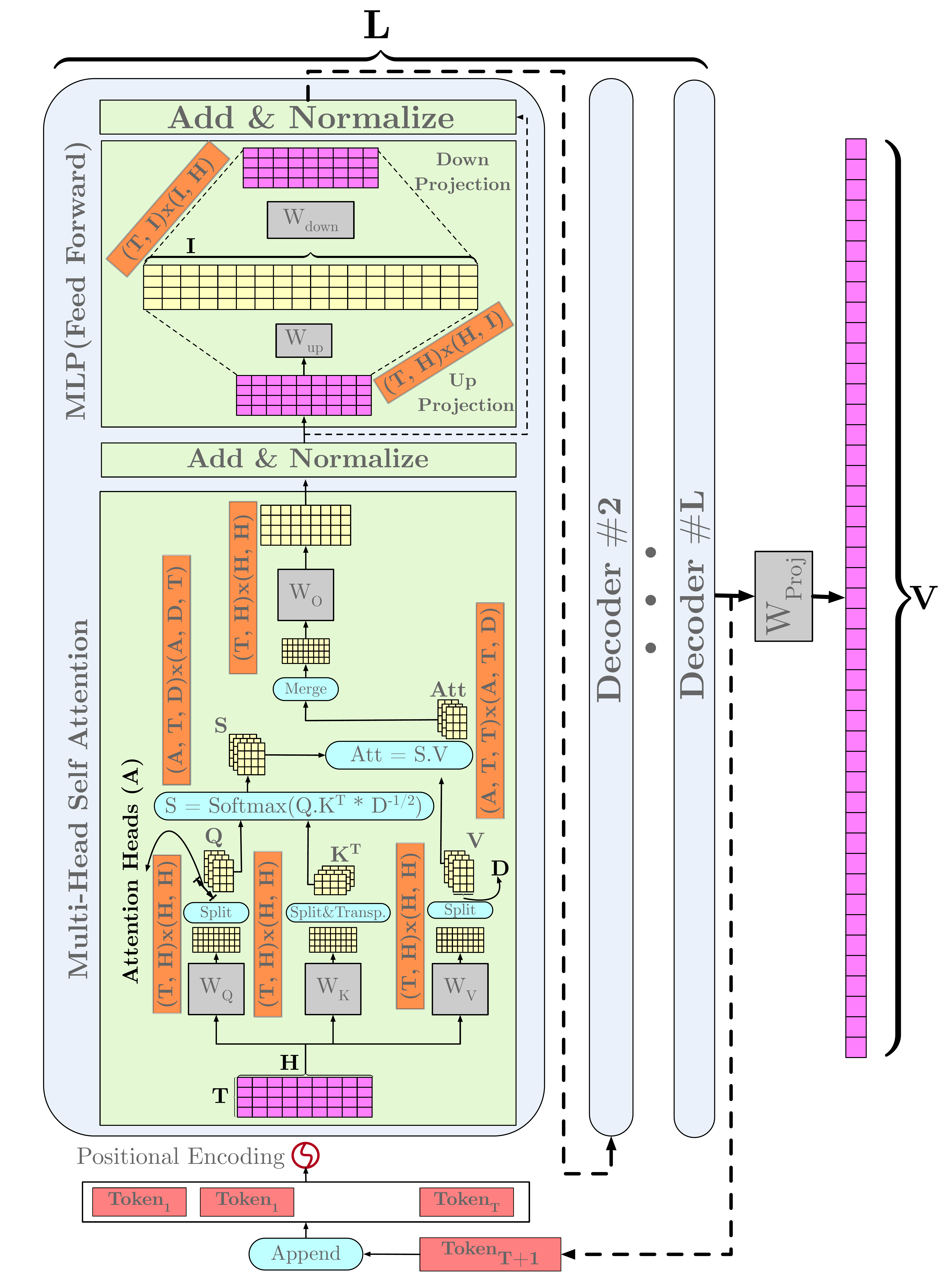}
    \caption{\textbf{The architecture of a decoder-only transformer. }\textit{The amount of computation (matmul dimensions) required for generation of each token depends on 5 key parameters: \(H\), \(L\), \(A\), \(I\), and \(T\). }}
    \label{fig:transformer_arch}
\end{figure}

The Transformer architecture~\cite{att_all_you_nead} is the foundation of nearly all modern LLMs, such as GPT~\cite{OpenAI_GPT5_SystemCard_2025}, Gemini~\cite{geminiteam2025geminifamilyhighlycapable}, and LLaMA~\cite{Touvron2023LLaMAOA}. Decoder-only variants~\cite{Radford2018ImprovingLU,radford2019language}, optimized for next-token prediction, are the basis of most generative LLMs today. As shown in Figure~\ref{fig:transformer_arch}, a decoder-only transformer consists of a closed loop of repeated decoder blocks---we denote their count as \(L\)---along with an \textit{Embedding} block at the input and a \textit{projection} block at the output. Each decoder block is composed of several subcomponents, the most prominent being \textit{Attention}, \textit{MLP}, and \textit{Normalization}. While numerous variants, particularly of the attention mechanism (e.g., GQA \cite{Ainslie2023GQATG}), have been introduced, their underlying computational structure remains largely similar.

The Transformer processes an input sequence of \(T\) tokens, each represented by a hidden vector of dimension \(H\), which serves as the primary dimension propagated through the network. The sequence matrix is projected into multiple attention heads---their count denoted by \(A\)---to compute attention scores, followed by an MLP subcomponent that expands the representation to a larger intermediate dimension--- \(I\)--- before projecting it back to \(H\). Normalization layers are interleaved between major components. The embedding and projection layers map the tokens between the one-hot token representation of length equal to the vocabulary size \(V\) (typically public) and the hidden dimension space.  

Overall, these five parameters---\(H\), \(L\), \(A\), \(I\), and \(T\)---dictate most computations within a decoder-only LLM. Among them, the first four (\(H, L, A, I\)) define the model’s {architectural configuration}, while the fifth (\(T\)) is externally controlled by the user through the {prompt length}. We refer to these parameters extensively throughout \S\ref{sec:attack2-leakarch}. 

Early implementations of attention on modern GPUs simply decomposed the computation into matrix multiplications executed using cuBLAS-based~\cite{NVIDIA_cuBLAS} GPU kernels.
Later works have focused on improving the efficiency of attention through optimized memory access patterns, operator fusion, and better utilization of modern GPU hardware features. A prominent example is the FlashAttention family of kernels~\cite{dao2022flashattentionfastmemoryefficientexact, dao2023flashattention2fasterattentionbetter}, which leverages on-chip SRAM as a software-managed cache to reduce memory traffic and compute the softmax operation more efficiently. 

In this work, we consider both implementations. We use a standard implementation based on native PyTorch linear algebra primitives provided by the HuggingFace \texttt{Transformers} library~\cite{wolf-etal-2020-transformers}, which we refer to as \textit{eager attention}. 
For optimized implementation, we use FlashAttention2 as a representative baseline in our evaluations in \S\ref{sec:attack2-leakarch}.

\subsection{Autoregressive Decoding}

The Transformer architecture generates one token per forward pass by computing a conditional probability distribution over the vocabulary, given the starting sequence of tokens for each pass. More precisely, for a sequence of tokens \(x_1, x_2, \ldots, x_T\) at the start of pass \(T+1\), a Transformer-based LLM predicts the next token \(x_{T+1}\) by estimating
$P(x_{T+1} \mid x_1, x_2, \ldots, x_T)$
and then sampling a token from the resulting distribution. The newly generated token is appended to the end of the current sequence, and the process repeats iteratively to produce subsequent tokens. This property also allows users to observe the model's output as a continuous stream of tokens, eliminating the need to wait for the entire response to be generated before accessing it.
Although this approach is conceptually simple and effective, it is computationally expensive. As model sizes continue to increase and larger context lengths are introduced, the latency of generating each token grows significantly, even taking several seconds per token for large models. 
As a result, a wide range of optimization techniques has been introduced to reduce token-generation latency~\cite{Jacob_2018_CVPR,Jo2025FastKVKC,dao2022flashattentionfastmemoryefficientexact}.

\subsection{Key-Value Caching}
Caching key–value (KV) pairs is a widely adopted optimization for improving inference efficiency~\cite{Jo2025FastKVKC}. During generation, the model reuses previously computed attention keys and values for past tokens, avoiding redundant recomputation at each step.
As a result, inference is typically divided into two stages: \textit{Prefill}, where the model processes the entire input prompt and constructs the KV cache, and \textit{Decoding}, where one token is generated per forward pass using the cached representations. 
The prefill stage is compute-intensive due to full-sequence processing, while the decoding stage is typically memory bandwidth–bound, as it incrementally attends to the growing KV cache to generate a single new token per step. This distinction leads to different runtime characteristics across the two stages, which we carefully account for in our modeling in \S\ref{sec:attack2-leakarch}.

\subsection{Speculative Decoding}
\label{sec:background:optimizations}

The observation that autoregressive models can verify subsequences of their input tokens in parallel with the next-token prediction~\cite{sternBlockDecode} has led to a new class of optimization techniques~\cite{Cai2024Medusa,Andrea2023Parallel,Fu2024Lookahead,Kou2024CLLM,raposo2024mixture}.
A prominent example is speculative decoding~\cite{leviathan2023spec, chen2023accelerating,xia-etal-2023-speculative,MiaoSpecInfer,sun2024spectrfastspeculativedecoding,BildSpec,chen2025cascadespeculativedraftingfaster,spector2023acceleratingllminferencestaged, bhendawade2024speculative,he-etal-2024-rest}, a technique inspired by speculative execution in processors~\cite{hennessy_patterson2017}. 
This optimization relies on a collaboration between two differently sized models: a \textit{main} target LLM, which is large and capable, and a \textit{draft} model, which is smaller, faster, and has modest generation quality. In each generation round, the draft model proposes a sequence of candidate tokens, which are then verified by the main model in a single forward pass. Depending on the level of agreement between the two models, multiple tokens may be accepted in one pass, reducing the number of expensive invocations of the main model in the average case. Thus, speculative decoding does not uniformly accelerate every generation, but it substantially improves the common-case inference performance. Crucially for this work, the interaction between the draft and main models introduces input-dependent timing variation, creating a timing side channel that we exploit to detect the optimization and recover its deployment details.

%-------------------------------------------------------------------------------
%%  Threat Model
%-------------------------------------------------------------------------------
%-------------------------------------------------------------------------------
\section{Threat Model}
%-------------------------------------------------------------------------------

\niparagraph{\textbf{Goals}.}
We assume an adversary who seeks to infer architectural and deployment information about a production large language model (LLM) by observing its per-token generation timing. 
The adversary’s goal is to recover information about any or all of the following key architectural parameters:  the number of decoder layers (L), hidden dimension size (H), number of attention heads (A), and the intermediate size of the MLP layer (I). 
Additionally, the adversary aims to identify inference-time optimization techniques employed by the provider, particularly the use of speculative decoding~\cite{leviathan2023spec} and the context length of the draft model used.

\niparagraph{\textbf{Target Model}.}
We assume the target model is a decoder-only~\cite{Radford2018ImprovingLU} transformer-based~\cite{att_all_you_nead} model accessible through a standard public web interface or API with streaming generation enabled. In this setting, the model transmits response tokens to the user as they are generated, rather than returning the complete output after generation is finished.

\niparagraph{\textbf{Capabilities}.}
The attacker can observe and measure the timing of the tokens received from the model API. The adversary operates entirely within the standard user interface exposed by the provider. We do not assume any additional control over request or generation parameters, nor access to internal model data such as logits, activations, or system logs. The attack also requires no privileged network position or system-level access.
The adversary has no prior knowledge of the target model’s internal architecture or deployment configuration beyond publicly available information, such as the model name, family, advertised capabilities, or maximum context length. 
For the model architecture leakage attack described in \S\ref{sec:attack2-leakarch}, we assume that the attacker knows the GPU type used to execute the target model and has access to timing data collected on the same GPU class to train the runtime predictors. The attack further assumes a single-GPU inference setting and does not model multi-GPU execution.

%-------------------------------------------------------------------------------
%%  Attack 1: Extracting Inference Optimizations
%-------------------------------------------------------------------------------
%-------------------------------------------------------------------------------
\section{Leaking Inference Optimizations}
\label{sec:attack1-infopt}
%-------------------------------------------------------------------------------

As large language models scale to billions of users and ever-larger architectures, inference has become increasingly compute-intensive and resource-constrained. 
Providers must balance model quality, latency, and energy efficiency to sustain acceptable Quality of Service (QoS) at global scale.
To achieve this, they aggressively optimize the inference pipeline through specialized hardware~\cite{grok}, custom serving infrastructure~\cite{park2025surveyinferenceengineslarge}, and proprietary inference-time techniques~\cite{park2025surveyinferenceengineslarge}.
These optimizations may be valuable intellectual property that gives leading providers a competitive advantage as the market expands and new entrants emerge~\cite{baseten_inference_platform}. 
Maintaining the confidentiality of these details is therefore a strategic priority, particularly for frontier models operated at massive scale.
In this section, we demonstrate that these proprietary deployment and optimization details are vulnerable to timing side channels. Specifically, we present an attack that shows an external observer can infer the presence of speculative decoding in production systems and further estimate key hyperparameters, including the context length of the draft model used in the speculative decoding pipeline.

\subsection{Attack Overview}
\label{sec:attack1-overview}

As discussed in Section~\ref{sec:background}, speculative decoding is an inference optimization technique inspired by speculative execution in modern processors~\cite{hennessy_patterson2017}. 
It aims to reduce end-to-end token-generation latency by using a small, fast draft model to speculatively predict N draft tokens, which are then verified in a single step by a larger target model~\cite{leviathan2023spec}. 
If the outputs of the draft and target models agree, all N tokens are accepted, and the draft model proceeds to generate the next speculative tokens. 
This process reduces the number of full forward passes of the large model by approximately a factor of N in the best case, when the smaller model’s predictions match with the larger model.
In contrast, if the target model disagrees with the draft, only one token (the large model’s next token) is accepted, and the remaining draft tokens are discarded, resulting in a worst-case latency comparable to that of standard non-speculative decoding.

Inspired by speculative execution vulnerabilities in modern processors~\cite{spectre}, we exploit the input-dependent timing variations that arise from disagreement between the draft and target models during speculative decoding.
The key idea is to intentionally induce controllable disagreement between the two models through carefully crafted prompts.
If such disagreement exists, it should cause observable slowdowns in per-token generation time, revealing the presence of speculative decoding in the inference pipeline.

To achieve this, we leverage the observation that in most deployments, the draft model used for speculation has a shorter context length than the main model. 
This design choice is intuitive: smaller models are optimized for speed and memory efficiency, and extending their context window significantly increases both computational cost and memory footprint.
We confirm this pattern using publicly available information on speculative decoding deployments including examples from \cite{long_context_spec2025}.

If a prompt requires a context longer than the draft model’s maximum context length, and the system employs speculative decoding, the draft model inevitably produces incorrect predictions.
The speculative tokens are then rejected, and the token-generation latency falls back to that of the large model. 
In contrast, when the prompt fits within the draft model’s context window and does not demand high model capacity, the draft model’s predictions are verified by the target model, resulting in faster generation.

By gradually increasing the required context length across prompts, we can observe where a sudden increase in per-token latency occurs.
The appearance of such a latency spike indicates that the provider is using speculative decoding. Furthermore, the position of this spike reveals the approximate context length of the underlying draft model.

\subsection{Crafting the Prompt}
We aim to craft a set of prompts that elicit responses with distinctive timing signatures capable of differentiating models that employ speculative decoding from those that do not use any inference-time optimization, as well as from models that implement alternative optimizations discussed in \S\ref{sec:background:optimizations}.

If we observe input-dependent variability in token-generation timing, it suggests the presence of some form of inference-time optimization. 
The more challenging task, however, is to craft prompts that specifically expose timing patterns unique to speculative decoding, rather than those caused by other inference-time optimizations.
A defining characteristic of speculative decoding is that it relies on a smaller draft model, which typically has a shorter context length than the main model, as discussed above. 
Therefore, when a response requires a context longer than the draft model’s context length, we expect a noticeable spike in generation time. 
To isolate this effect and ensure that timing variability is indeed caused by context length, we design prompt sets where context length is the only variable, and all prompts require comparable (if not equal) semantic complexity from the model.

To that end, we select a simple task of remembering a large numeric value presented in the early part of the prompt. Concretely, we use the following template:

\medskip\noindent
\resizebox{\linewidth}{!}{
  \centering \texttt{
    $\begin{array}{l}
      \overbrace{\text{We have a }\{n\} \text{ digit number NUM=}}^{prologue\_length}\{rand\_num\}_n. \{var\_pad\_str\}_x \\
      \text{The value of number NUM at}
      \text{ the start was equal to}\end{array}$
}}

\leftskip=0pt\medskip
Here $\{s\}_n$ denotes a string $s$ with $n$ characters. String $\{rand\_num\}_n$ is a uniformly sampled $n$-digit integer expressed with ASCII digits, and $\{var\_pad\_str\}_x$  is a sequence of $x$ randomly sampled alphabetical characters used to reach a desired prompt length. Additionally, $prologue\_length$ is the length of the beginning part of the prompt before the generated random number.

We avoid fixed numbers or static padding to reduce the chance that other server-side optimizations (e.g., caching) influence timings. By gradually increasing the padding size $x$, we sample prompt lengths up to the reported maximum context length of the victim API. As we increase $x$, the context length required to generate a correct response increases. For each attack iteration, we generate a fresh set of prompts of varying lengths from this template. We randomize the order in which prompts of different lengths are sent to the victim API to rule out the effects of potential order-dependent or prompt-length–dependent throttling policies on the observed timing behavior.

\subsection{Detection and Parameter Estimation}

We perform a two-stage inference procedure to identify and characterize the use of speculative decoding in the victim LLM:

  \niparagraph{\textbf{Detection.}} For each sample prompt, we plot the per-token generation timing of the victim LLM. A detectable timing spike after a certain input length indicates data-dependent behavior that depends solely on prompt length rather than semantic complexity. Such a pattern suggests that the victim employs speculative decoding. 
  Beyond a certain input length, the smaller draft model can no longer generate accurate speculations, causing the generation to be predominantly handled by the large target model, with additional overhead from the execution of the draft model.
If no timing spike is observed, the presence of speculative decoding cannot be ruled out, as it remains possible, though less likely (\S\ref{sec:attack1-overview}), that the draft model operates with the same context length as the main model. Consequently, this detection method may yield false negatives.
  
  \niparagraph{\textbf{Parameter Estimation.}} Upon detecting speculative decoding, we can further infer the context length of the draft model. Specifically, we identify the critical prompt length at which the timing deviation occurs by performing a binary search within the region of the timing spike. The middle point of each search iteration is tested until the exact breakpoint is located ($T_{break}$). Using this value, we estimate the context window length of the smaller draft model as  ($ T_{break} - prologue\_length$), which is the difference between the breakpoint and the number of tokens in the prompt before the first token required to produce the correct answer (i.e., the first character of $\{rand\_num\}_n$).

\subsection{Experimental Setup}
\label{sec:attack1-setup}

We conduct two sets of experiments: one using a local implementation of speculative decoding to validate our detection approach, and another using black-box access to unknown production models to demonstrate its applicability in the wild. Below, we describe the experimental setup for each case.

\subsubsection{Local Experiments}
To characterize the timing behavior of speculative decoding, we use a publicly available third-party implementation~\cite{LLMSpeculativeSampling_2025} of the speculative decoding technique originally proposed by Leviathan et al.~\cite{leviathan2023spec}.

We use Guanaco 13B (a fine-tuned LoRA layer for LLaMA2 13B)~\cite{guanaco13b_dettmers_2025} as the main target model and TinyLLaMA 1.1B~\cite{TinyLlama1.1B_2025} as the draft model. 
All experiments are performed on a cloud-based system equipped with an NVIDIA L40 GPU with 45 GB of memory, running PyTorch 2.8 and Hugging Face Transformers 4.56 on Ubuntu 22.04.
We instrument the inference code with Python’s \texttt{time} library to record timestamps immediately before and after each generation round, enabling the computation of per-token generation timings.

\subsubsection{Remote Experiments}
\label{sec:attack1-remote-setup}

We repeat the same attack against several major publicly available LLM APIs, including Google Gemini, OpenAI GPT, Cohere, and Mistral models. Because the attack requires varying prompt lengths up to provider context limits (which can reach up to 1M tokens), we adjust the number of repetitions in the remote experiments to limit the cost. 
We use the streamed chat-completion interface for all experiments.
The client runs on an Ubuntu 22.04 host. To capture per-token timing, we record a timestamp at the arrival of every stream event and log the associated token count without additional on-the-fly processing. %on-the-fly
An overview of the experiment specifications is presented in Table~\ref{table:attack1_exp_setup}. A detailed analysis of the effect of the network on our measurements can be found in appendix \S\ref{appendix-spec-net}.

    % \begin{table}[t]
    % \centering
    % \caption{Specifications of the Inference Optimization Extraction attack}
    % \begin{tabular}{l c c c c} 
    %  \toprule
    %  \textbf{Parameter} & \textbf{Llama} & \textbf{Gemini} & \textbf{GPT} & \textbf{Mistral}\\ 
    %  \midrule
    % Prompt Length (Tokens) & [10, 4k] & [2k, 1M] & [1k, 128k] & [1k, 128k]\\ 
     
    % \#Trials & 100 & 30 & 50 & 40\\ 
    
    % Open Model? & \cmark & \xmark & \xmark & \xmark \\
    
    % Local? & \cmark & \xmark & \xmark & \xmark \\
    
    % Cost/1M Input Tokens & 0\$ & 0.075-0.30\$ & {\color{red} 0.50-10\$} & 2\$ \\
     
    %  \bottomrule
    % \end{tabular}
    % \label{table:attack1_exp_setup}
    % \end{table}

\begin{table}[t]
\small
\centering
\caption{Specifications of the inference optimization extraction experiments.}
\label{table:attack1_exp_setup}
\begin{tabularx}{\columnwidth}{@{} l *{3}{>{\centering\arraybackslash}X} @{}}
  \toprule
  \textbf{Parameter} & \textbf{LLaMA13B} & \textbf{Gemini} & \textbf{Cohere} \\  %\textbf{OpenAI GPT} & \textbf{Mistral}\\ 
  \midrule
  % Prompt Length (Tokens) & 10-4K & 2K-1M & 1K-128K & 1K-128K \\
  Context Length & 4K & 1M & 128K \\ % 128K & 128K \\
  \# Trials & 100 & 30 & 65 \\ %50 & 40\\
  Open Model? & \cmark & \xmark & \cmark \\ % \xmark & \xmark \\
  Local? & \cmark & \xmark & \xmark \\ % \xmark & \xmark \\
   \$ / 1M Tokens & 0 & 0.3 & 2.5 \\ % 0.5--30 & 2.0 \\
  % Avg RTT (ms) & \_ & 26.65 & 5.75 \\ % 2.066 & 2.010 \\
  % RTT std dev (ms) & \_ & 0.07 & 0.08 \\ % 0.260 & 0.121 \\
  \bottomrule
\end{tabularx}
\end{table}

\subsection{Results}
\label{sec:attack1-results}

\subsubsection{Local White-Box Model Experiment}

We first consider using a locally deployed model that employs speculative decoding to validate our detection technique.

We generate prompts from our prompt template above with prompt length from 82 to 4K and we observe per-token generation times. If a generation event produces multiple tokens simultaneously, we divide the total per-event generation time by the number of tokens to approximate the per-token generation time. 

\begin{figure}[t]
  \centering
  \includegraphics[width=0.85\columnwidth]{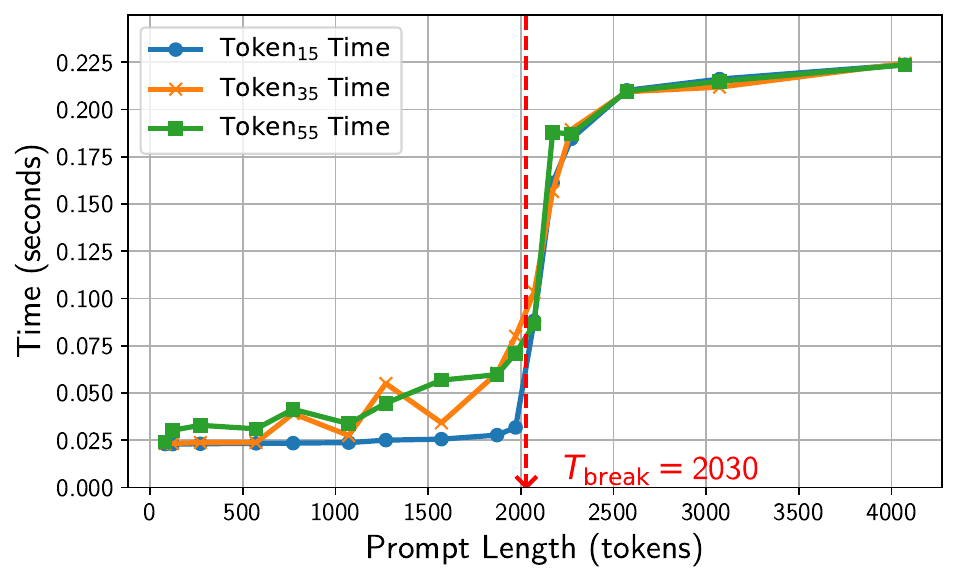}
  \caption{\textbf{Timing pattern for a local implementation of speculative decoding. }\textit{After a specific prompt length, the small draft model cannot produce useful predictions due to its smaller context length, causing a noticeable spike in timing.}}
  \label{fig:attack1-results-local-detect}
\end{figure}

Figure~\ref{fig:attack1-results-local-detect} presents the results of this experiment. We observe a sharp rise in per-token generation time at a specific input length, consistent with the expected behavior of speculative decoding. 
In a separate experiment, we observe that disabling speculative decoding eliminates this spike, confirming that the effect originates from the speculative decoding mechanism.

We next evaluate whether the context length of the draft model, TinyLLaMA 1.1B, can be inferred solely from per-token timing measurements. We perform a binary search over input prompt lengths to identify the breakpoint where token generation timing shifts from optimal efficiency to the worst-case scenario. Using the same prompt template, we generate prompts spanning the two ends of the timing jump observed in Figure~\ref{fig:attack1-results-local-detect}.

After narrowing the search, we observe that the timing jump occurs approximately at $T_{break} = 2030$ tokens. Applying our context estimation formula to this value yields an estimated context length of 2018 tokens for the draft model. Given that model context lengths are typically of powers of two, we refine our estimate to the nearest power-of-two value, 2048 tokens, as the effective context window of the draft model. This matches the context length of TinyLLaMA 1.1B, validating the accuracy of our parameter retrieval method.

\subsubsection{Remote Black-Box Model Experiments}
We next evaluate our attack on popular publicly available LLM APIs (listed in \S\ref{appendix-spec-A}) in a black-box setting.

\niparagraph{\textbf{Detection Results.}} We repeat the same tests from the local experiments on the chat-completion APIs of the target providers. After extracting and processing the per-token generation timings, we plot the results for each remote LLM using the same methodology described earlier. Several models exhibit the timing spike associated with speculative decoding, while others show no such evidence.

\begin{figure}[t]
  \centering

  \begin{subfigure}[t]{0.49\columnwidth}
      \centering
      \includegraphics[width=\textwidth]{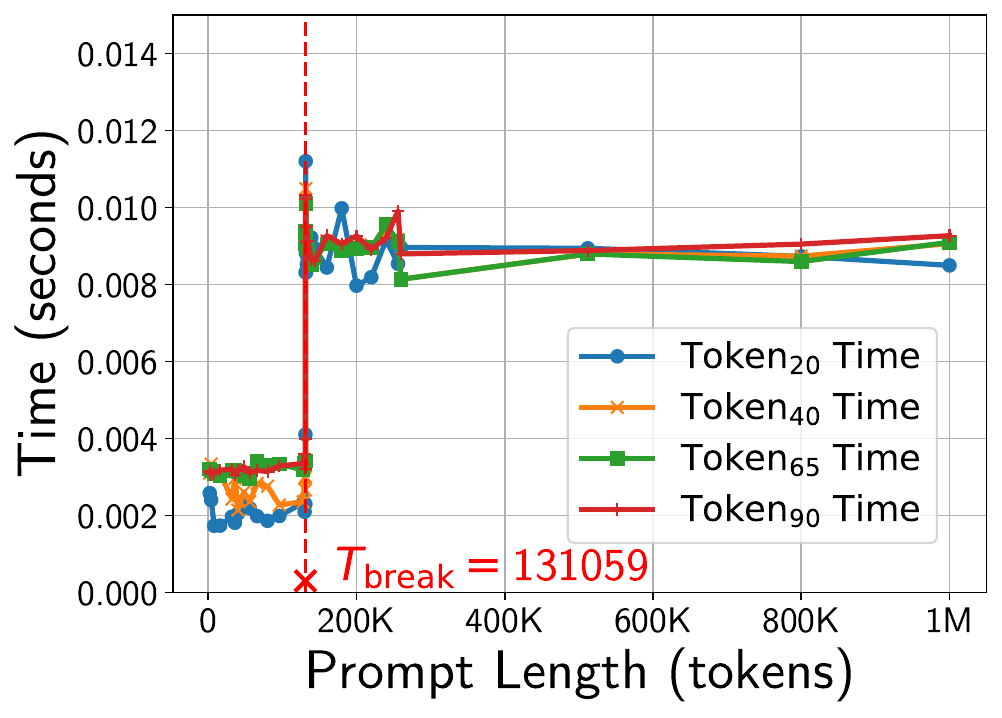}
      \caption{Gemini Flash-2.5}
      \label{fig:attack1-results-remote-flash25}
  \end{subfigure}
  \hfill
  \begin{subfigure}[t]{0.49\columnwidth}
      \centering
      \includegraphics[width=\textwidth]{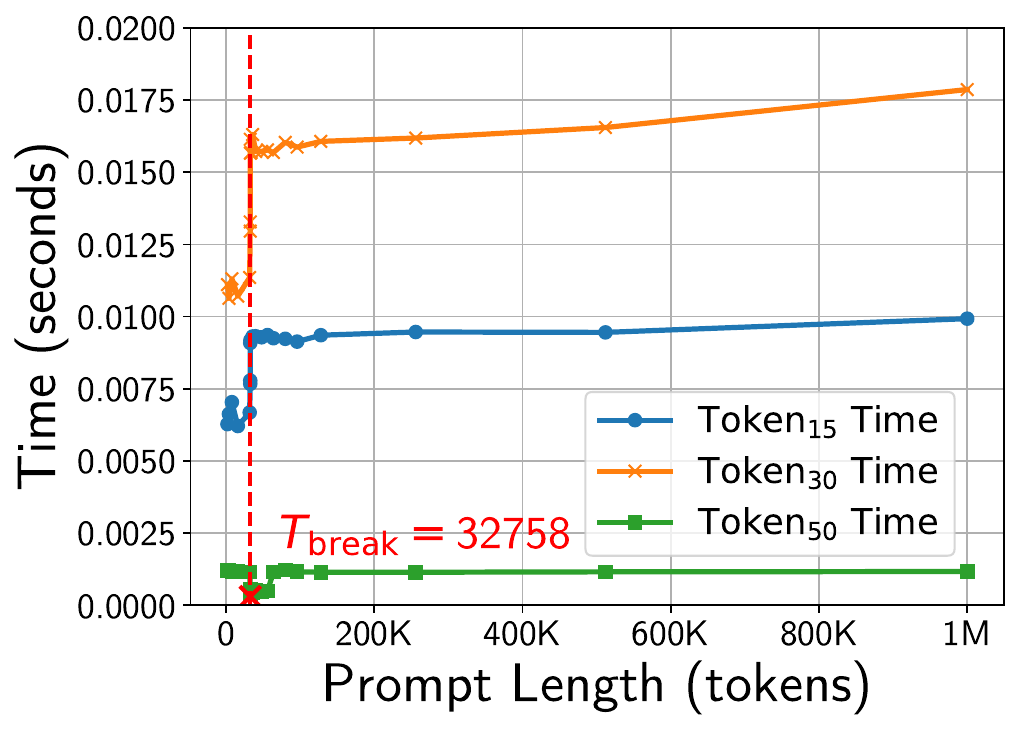}
      \caption{Gemini Flash-1.5}
      \label{fig:attack1-results-remote-flash15}
  \end{subfigure}

  \caption{\textbf{Per-token generation time of Gemini models.} \textit{These models exhibit speculative decoding behavior. Flash-2.5-Lite also shows a similar timing jump as depicted in \S\ref{appendix-spec-A}}.}
  \label{fig:attack1-results-remote-success}
\end{figure}

Figure~\ref{fig:attack1-results-remote-success} presents the per-token generation time for different input prompt lengths for Gemini Flash-2.5 and Gemini Flash-1.5. Both exhibit a distinct spike in the timing curve at a specific prompt length, an indicator of speculative decoding. The Flash-2.5 and Flash-2.5-Lite models are among the most recent and widely used models in Google AI Studio. 

In contrast, for several other APIs, we do not observe any such timing pattern. While this absence of detectable signatures prevents us from confirming the use of speculative decoding, it does not preclude its presence as our approach can produce false negatives, as described earlier.

\niparagraph{\textbf{Parameter Estimation.}} For the Gemini models in which we detect the use of speculative decoding, we perform a binary search similar to that described in the local experiment to estimate the draft model’s context window length. The extracted values are summarized in Table~\ref{table:attack1_results_hyperparam}.

\begin{table}[t]
\small
\centering
\caption{Retrieved context window lengths of the draft models in speculative decoding–enabled LLMs.}
\label{table:attack1_results_hyperparam}
\begin{tabularx}{\columnwidth}{@{} l *{1}{>{\centering\arraybackslash}X} @{}}
  \toprule
  \textbf{LLM Model} & \textbf{Draft Context Length} \\
  \midrule
  Guanaco 13B + TinyLlama 1.1B & 2{,}018 $\rightarrow$ 2K \\
  Gemini Flash 2.5 & 131{,}042 $\rightarrow$ 128K \\
  Gemini Flash 2.5 Lite & 131{,}042 $\rightarrow$ 128K \\
  Gemini Flash 1.5 & 32{,}743 $\rightarrow$ 32K \\
  \bottomrule
\end{tabularx}
\end{table}

%-------------------------------------------------------------------------------
%%  Attack 2: Leaking Model Architecture
%-------------------------------------------------------------------------------
%-------------------------------------------------------------------------------
\section{Leaking Model Architecture}
\label{sec:attack2-leakarch}
%-------------------------------------------------------------------------------

The architecture and parameter configurations of leading LLMs are considered highly confidential, both for competitive and strategic reasons. 
We show that timing leakage from publicly accessible models can serve as a powerful source of information about these systems. 
To demonstrate this, we present an attack capable of recovering key architectural details, such as the hidden dimension size and number of decoder layers, among others. 
To achieve this, we build an accurate timing model of the LLM inference pipeline that captures how generation latency varies with different model properties, and we use this model to reverse-engineer unknown architectures.

This section describes the attack, the construction of the timing model, and how it enables inferring parameters of black-box LLMs.

\begin{figure}
    \centering
    \includegraphics[width=\linewidth]{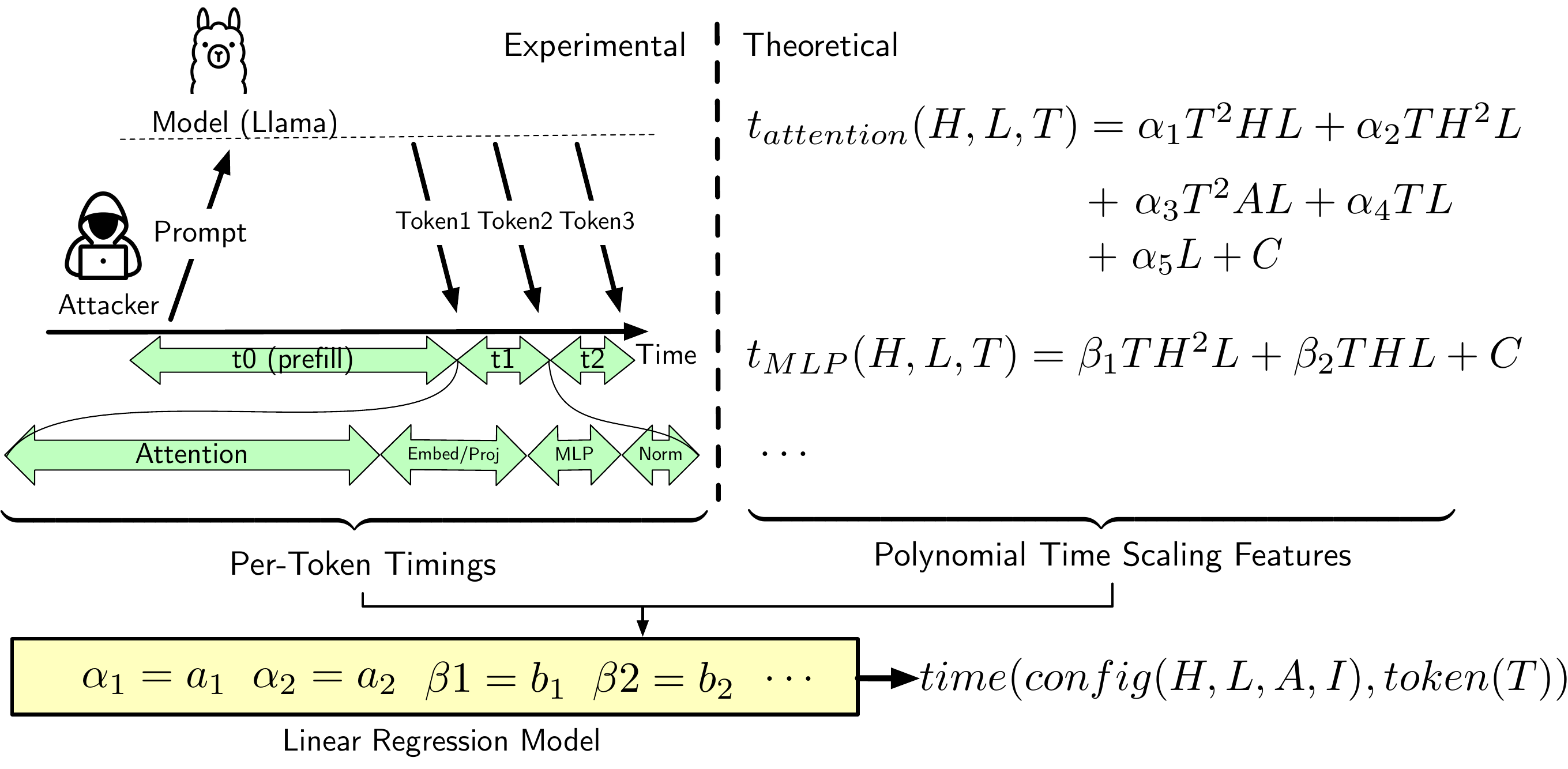}
    \caption{\textbf{Overview of the process for constructing the LLM runtime predictor model. }\textit{Empirical data is used to learn coefficients of asymptotic terms derived analytically.}}
    \label{fig:attack2-overview-overall-predictor-workflow}
\end{figure}

\subsection{Attack Overview}
\label{sec:attack2-overview}

A key component of this attack is constructing a model that captures the relationship between per-token generation time and the model parameters we aim to recover, namely, the number of attention heads (A), hidden dimension size (H), number of decoder layers (L), and intermediate size (I). 
However, accurately modeling computations as complex as transformer inference on modern GPUs is highly challenging. To address this, we adopt a bottom-up framework that begins with theoretical asymptotic analysis and progressively incorporates empirical corrections to capture hardware- and implementation-specific effects.

We leverage the modular structure of the transformer architecture by decomposing its runtime into primitive operations and incrementally increasing modeling complexity by composing these primitives into higher-level components.
We perform asymptotic analysis on each primitive to characterize how its runtime scales with respect to input parameters, expressed in terms of computational and memory costs. 
Next, we introduce multiplicative coefficients and constant terms to capture additional factors influencing runtime.
These coefficients are learned from empirical measurements, allowing the model to account for hardware- and implementation-specific effects.

We instrument real implementations of transformer components, measure their runtimes across a range of input sizes and architectural dimensions, and fit regression models using the asymptotic cost terms as features. The learned coefficients capture hardware- and implementation-specific effects. Once calibrated, this hybrid analytical–empirical model can estimate per-token runtime for new architectures and input sequences with good accuracy. The overall workflow is shown in Figure~\ref{fig:attack2-overview-overall-predictor-workflow}.

Although many approaches could be used to predict runtime, including large neural networks, we deliberately adopt simple linear regression. Linear models provide three key advantages: they are explainable, since their terms directly correspond to theoretical cost components; generalizable, as they are less prone to overfitting across diverse architectures; and configurable, allowing easy adjustment or extension of runtime terms.

We then use the predictor as an oracle to infer architecture for a black-box model. Given a target, we control input prompts and collect per-token timing traces. We then construct a search space of plausible LLM configurations (e.g., H, A, L, I). To avoid explosive search costs, we prune the space using practical observations (e.g., typical alignment multiples such as 64 or 128). With a feasible search grid, we enumerate plausible configurations, synthesize timing traces for each candidate using the predictor, and rank candidates by distance to the observed trace. The top matches are the most likely architectural configurations for the target model.

Different implementations of the transformer architecture exhibit different asymptotic timing behavior. 
These implementations may rely on different GPU kernels (e.g., FlashAttention vs. cuBLAS kernels) or incorporate optimizations such as KV-cache to avoid recomputation.
We begin with a baseline eager attention implementation~\cite{att_all_you_nead,wolf-etal-2020-transformers} that decomposes attention into matrix multiplications executed via cuBLAS. We then extend the analysis to more advanced implementations, such as FlashAttention, and finally model the impact of KV-cache optimization.

This approach assumes that once characterized, an implementation's asymptotic behavior and the associated coefficients remain invariant across input sizes, allowing runtimes to be extrapolated to unseen dimensions.
While this assumption generally holds at the level of individual kernels, CUDA libraries such as cuBLAS (used only in eager implementation) dynamically select from a large set of specialized kernels based on operand shapes and internal heuristics. This leads to small but non-negligible deviations from simple scaling terms.
To address this, we model kernel selection and per-kernel runtimes in cuBLAS for matrix multiplications and apply online corrections for affected operations and components, improving prediction accuracy for unseen configurations.

This attack shows that timing traces carry an architecture-dependent fingerprint that is not easily reproduced by different architectures. In other words, if the per-token runtime is viewed as a quadratic function of input length, the coefficients of that function are themselves determined by a polynomial combination of the model’s architectural parameters. Different architectures, therefore, yield different coefficient sets. As a result, with sufficiently precise timing observations across a diverse set of inputs, an adversary can reliably infer an approximate internal architecture of a black-box LLM. Moreover, incorporating results from other attacks that leak specific architectural parameters, such as the hidden dimension size \cite{carliniStealProd}, can further reduce the size of our search space and improve the accuracy of the remaining inferred parameters.

\subsection{Offline Phase}

We first construct a per-token runtime predictor by instrumenting and profiling a set of open-access LLM implementations. The collected measurements are analyzed to build an analytical–empirical model that maps architectural parameters and input dimensions to per-token runtime estimates. In addition, we develop a matmul runtime corrector model to address the non-linearities that arise when handling unseen architectural dimensions during the online phase.

\niparagraph{\textbf{Building Polynomial Time–Scaling Features}.}
\label{sec:attack2-method-features}
We decompose each decoder block from a decoder-only LLM (shown in Figure~\ref{fig:transformer_arch}) with \(L\) decoder layers into its canonical subcomponents (multi-head attention, MLP, normalization, etc.) and enumerate the primitive algebraic operations performed by each subcomponent (matrix multiplications, softmax, element-wise additions, scalings, \(\ldots\)). For each primitive operation, we derive its theoretical compute and memory cost as a function of the relevant dimensions, including the sequence length \(T\), hidden size \(H\), number of attention heads \(A\), MLP intermediate projection dimension \(I\), and bit-width of each element in bytes \(b\) (e.g., 2 for fp16 or 0.5 for int4).

As a concrete example, consider a single decoder block with a naive Q-projection that multiplies a token embedding matrix of size \(T\times H\) by a weight matrix \(W_Q\) of size \(H\times H\). 
A naive multiplication algorithm requires $O(TH^2)$ operations and $O(bTH+bH^2)$ memory reads.

To convert these theoretical costs to wall‑clock time, we use the platform characteristics: the maximum rate by which we can execute arithmetic operations, i.e., peak arithmetic throughput \(d\) (ops/sec), and the rate by which we can fetch data from memory, i.e., memory bandwidth \(c\) (bytes/sec). The baseline runtime contribution of a cost term is therefore the cost divided by the corresponding rate (e.g., $O(TH^2)/d$ and $O(bTH+bH^2)/c$ for the Q-projection example).

While different implementations may, in theory, exhibit varying asymptotic behavior, we find that in practice these naive expressions provide good approximations.
Hardware and software optimizations such as tiling, vectorization, and kernel launch overhead primarily affect the empirical coefficients of these terms rather than their asymptotic dependence on
our parameters of interest (\(T, H,\ldots\)). We therefore express the runtime of a subcomponent as a linear combination of theoretical terms with empirical coefficients. For example, we can express the timing of our Q-projection example as

\begin{equation}
    time_{Q-proj} \approx \alpha \frac{TH^2}{d} + \beta\frac{bTH+bH^2}{c} + C
\end{equation}
where $\alpha,\beta,$ and $C$ are fitted coefficients that capture implementation and hardware overheads. In some implementations, memory costs can be completely hidden by computation when the two operations overlap. Although our model adds these terms linearly, this behavior is captured by the empirically learned coefficients. For example, the memory coefficient becomes effectively zero when memory latency is fully masked by computation.

Following the same process used for the single Q-projection matrix multiplication, we extend the analysis to a multi-head attention block with \(A\) heads and per-head dimension \(D = H/A\). We derive the theoretical scaling features for all major operations in this subcomponent. A complete list of the derived operations for each subcomponent is provided in the appendix~\S\ref{sec:appendix-A-1}.

Next, we move one level higher and compose the overall attention block scaling terms by combining the terms from operations of its subcomponents. To generalize the runtime of the attention block across the entire model, we multiply these scaling terms by the number of decoder layers \(L\), reflecting their repetition throughout the LLM.

\begin{equation}
    time_{attention} \sim T^2 HL + T H^2L + T^2AL +\;\ldots
\end{equation}
Following this example for the attention block, we perform a similar detailed analysis for all other subcomponents of the Transformer and extract their corresponding runtime scaling features (appendix~\S\ref{sec:appendix-A-2}).

\niparagraph{\textbf{Collecting per-Component per-Token Timings.}} To learn the empirical coefficients, we first collect detailed timing information from the actual implementation of the operations and subcomponents. We instrument an open-access model inference framework with timing utilities and record the runtime behavior of a reference model (e.g., Meta's Llama) for further analysis. We use a broad range of input prompt lengths (\(T\)) to probe the reference model and collect per-token timing data. These records serve as our reference dataset to estimate the runtime contribution of different terms and subcomponents.

To enrich the dataset, we generate multiple architectural variants of the same reference model by altering the number of layers (\(L\)) or the hidden dimension (\(H\)). This effectively produces many instances within the same model family, allowing us to capture diverse timing fingerprints without requiring access to multiple distinct models.

We construct a dataset that records the architectural properties of the dissected LLMs, input prompt lengths, and measured per-component per-token timings for each configuration, which serve as ground-truth labels. This dataset is then used in the next step of the offline phase to determine the appropriate coefficients for each term in the final runtime model of every component.

\niparagraph{\textbf{Composing the Predictor Model.}} Since the operations within a subcomponent follow a deterministic and predefined sequence, and the input length and architectural dimensions only affect the runtime of these operations rather than their execution flow, we can model the total runtime of a subcomponent as a simple linear combination of its operations.

We build a separate linear regression model for each subcomponent, which, given the input length and architectural dimensions of a target LLM, predicts the runtime of that specific component. Using separate regressors for individual subcomponents, in line with our bottom-up modeling approach, helps us verify the correctness of each part before aggregating them into the overall predictor. This modular verification process avoids the complexity of validating a large monolithic model and simplifies the integration of later corrections for specific components if needed. The total per-token generation time of the LLM is then obtained by summing the predicted runtimes of all components, scaled by the number of their instances within each transformer layer.

This ensemble of linear models already captures the dominant runtime scaling behavior of the architecture. Since we build the model component-wise and do not fit it directly to the runtime of the entire LLM, each subcomponent only captures the overheads within its own scope. To account for remaining latency sources outside the canonical components, such as kernel launch delays, synchronization costs, and constant per-iteration overheads, we include an additional linear regression component to model the residual overhead, ensuring comprehensive coverage of all timing contributions. We intentionally avoid using non-linear models such as radial basis function (RBF) regressors or neural networks, as they tend to overfit to residual noise and give a misleading impression of higher precision. Ultimately, we obtain linear equations that accurately approximate the runtime of each subcomponent. For example, for the attention block, we have
\begin{equation}
    t_{att} = a_{1} H^2TLd + a_{2} T^2HLd + a_{3} bT^2ALc + \ldots + C 
\end{equation}
where $a_1, \ldots, a_4$ and $C$ are the coefficients learned by the linear regression model on the runtime dataset. And for the entire model, we approximate the runtime of the entire model as 

\begin{equation}
    t_{llm} = t_{att} + t_{mlp} + 2 \cdot t_{norm} + t_{embed\_proj} + t_{overhead}
\end{equation}

where $t_{att}$, $t_{mlp}$, $t_{norm}$, $t_{embed\_proj}$, and $t_{overhead}$ denote the runtime of each of the components.

\niparagraph{\textbf{Training the Runtime Corrector}.}
To address the non-linearities in our runtime predictions caused by dynamic kernel selection in the underlying libraries, we prepare two auxiliary components: one that predicts which kernel the NVIDIA cuBLAS library~\cite{NVIDIA_cuBLAS} uses for each matmul size (i.e., pair of operand dimensions), and another that predicts the corresponding matmul runtime given its size and the chosen kernel.  

To train these components, we generate a comprehensive set of matmul operations representative of those invoked by an LLM across a wide range of configurations and profile both their runtimes and the executed kernel names. Using this dataset, we construct a two-phase ensemble model consisting of (1) a LightGBM classifier~\cite{lightGBM} that predicts the kernel selected for a given pair of operand shapes, and (2) a collection of small random forest regressors, each modeling the runtime behavior of a specific kernel based on empirical timing data. This two-component model later serves as part of our runtime correction logic, enabling our predictor to generalize effectively to all unseen configurations.

\niparagraph{\textbf{Modeling FlashAttention}.}
To demonstrate that our prediction framework generalizes to alternative Transformer implementations with minimal changes, we construct a new runtime predictor based on the FlashAttention2 kernel. We modify the runtime scaling terms associated with the attention subcomponent to reflect the theoretical behavior of FlashAttention2~\cite{dao2023flashattention2fasterattentionbetter}, and train a new predictor, \texttt{Flash2}, using the same methodology. The primary modification arises from the reduced memory access complexity of FlashAttention, which achieves an average-case of $\Theta(T^2 D^2 M^{-1})$ memory accesses, where $T$, $D$, and $M$ denote the sequence length, attention head dimension, and per-SM SRAM size, respectively. To capture this behavior, we introduce an additional memory-related scaling term into our attention model. These adjustments are derived from the algorithmic specification of FlashAttention rather than any specific implementation, and are listed in~\S\ref{sec:appendix-flash}. The runtime scaling terms for the remaining subcomponents (e.g., MLP) remain unchanged from the \texttt{Eager} predictor and are retrained using the new timing data. Finally, since FlashAttention2 does not rely on cuBLAS-based general matrix multiplication kernels, the matmul correction logic is not required for this predictor.

\niparagraph{\textbf{Modeling KV-Cache}.}
To further extend our framework to realistic autoregressive inference settings, we incorporate KV caching as a representative inference optimization into our runtime modeling. We construct a KV-cache-enabled predictor, \texttt{KV-Flash2}, by extending the \texttt{Flash2} model with separate runtime scaling terms for the prefill and decoding phases. The prefill stage follows the same scaling terms as the \texttt{Flash2} predictor, while the decoding stage terms (listed in~\S\ref{sec:appendix-flash-kv}) reflect single-token processing with attention computed over the cached sequence. This separation captures the change in computational complexity induced by KV reuse after the initial forward pass. The prefill regressors are trained using only the timings from the first forward pass of each prompt in the dataset, while the decoding regressors are trained on the timing data from the remaining passes. The final model is therefore a hybrid predictor that combines prefill and decoding formulations, enabling accurate modeling of long-context autoregressive inference workloads.

\subsection{Online Phase}
\label{sec:attack2-online-phase}

\niparagraph{\textbf{Building the Search Space.}} After the LLM runtime predictor is calibrated, we use it as an oracle. Given an architectural configuration and an input length, it returns a per-token runtime prediction for the output sequence. We then employ this oracle to search across the possible architectural configuration space, identifying configurations whose synthesized timing traces best match the observed timing sequence of a black-box target. To make this search tractable, we exploit several key practical observations about modern LLM design.

First, architectural dimensions are integer values, and this immediately discretizes the search space. Second, valid configurations follow implementation conventions and alignment constraints rather than arbitrary integer values. For example, in contemporary models, the number of decoder layers is typically an even integer, hidden dimensions are usually multiples of a base stride (e.g., 64 or 128), and the number of attention heads is chosen such that the per-head dimension divides the hidden size evenly. In addition, some combinations of parameters are inherently implausible, such as configurations with extremely small hidden sizes and very deep layer counts, or the reverse, which can be safely pruned to further reduce the density of plausible configurations.

Leveraging these constraints, we construct a multi-dimensional, discrete search grid over the unknown architectural parameters (e.g., \(L, H, A, I\)). 

\niparagraph{\textbf{Reverse Engineering the Target Architecture}.} With the search grid prepared and the predictor model serving as our oracle, we evaluate each candidate configuration using the same set of prompts that were employed to collect the timing traces from the target model. For every configuration–prompt pair, we query the predictor to generate the corresponding predicted timing sequence.

Next, we compute the distance between each predicted sequence and the observed target timings using a simple metric such as the root mean square error (RMSE). We then rank all candidate configurations according to this distance and select the top-ranked ones as potential matches for the target model’s architectural parameters.

Given that timing measurements are inherently noisy and that distinct configurations may occasionally yield similar timing profiles when probed with a limited set of prompts, we avoid over-reliance on the single closest match. Instead, we produce a short list of the nearest candidates for subsequent inspection, allowing for a small margin of error in the architectural inference.

\niparagraph{\textbf{Scaling Runtime for Unseen Dimensions}.}
\label{sec:online-runtime-correction}
When using the \texttt{Eager} predictor to estimate the runtime of configurations not observed during training, we incorporate the two corrector models trained in the offline phase. We first use the lightGBM classifier to predict the cuBLAS kernel that will be selected for each matrix multiplication in the target configuration. We then estimate the runtime of each matmul using the corresponding kernel-specific random forest regressor. These predictions are used to adjust the final runtime estimate of the \texttt{Eager} predictor for unseen architectural configurations.  We refer to this enhanced model as \texttt{Eager(Corrected)}. A detailed step-by-step description of this correction procedure is presented in Appendix~\S\ref{sec:app-online-runtime-correction}.

\niparagraph{\textbf{Ranking Strategy for Reverse Engineering with \texttt{KV-Flash2}}.}

Since we rely on the first output token to characterize the prefill phase, and the timing differences across subsequent decoding steps are minimal, we use the average latency of the remaining output tokens as a robust estimate of per-token decoding time for each prompt length. As a result, we obtain fewer effective data points compared to the earlier experiments, where each output token provided a strong timing signal. To address this, we leverage the hybrid design of the \texttt{KV-Flash2} predictor in a two-stage ranking procedure. We first narrow the search space to the top-35 candidate configurations using only the prefill component, and then refine the ranking within this subset by incorporating both prefill and decoding predictions. This staged approach improves the ranking of the true configuration and preserves the effectiveness of the attack under KV-cache–enabled inference.

\subsection{Experimental Setup}
\label{sec:attack2-exp-setup}

We use two models from the Llama~3.2 family with different sizes: the 1B version serves as our reference model for the offline phase to generate timing data for training the linear regression predictors, while the 3B version is used as the target model to generate test datasets for evaluating runtime prediction accuracy and the architectural search and reverse-engineering procedure. Unless stated otherwise, training (reference) datasets are generated from Llama~3.2~1B, and test (target) datasets are generated from Llama~3.2~3B.

The search grid for the reverse engineering task enumerates candidate values for each target architectural parameter, including but not limited to those observed in the datasets. The grid is strictly inclusive and is constructed to be at least one order of magnitude larger than the number of unique configurations in each dataset. The online search evaluates 1{,}540 candidate configurations for each of the 130 target configurations, and takes approximately 20 minutes when executed in parallel across 40 CPU cores. We report top-5 accuracy, defined as the fraction of target configurations for which the correct configuration appears within the top five candidates returned by the attack. A retrieval is considered successful if all targeted architectural parameters of a candidate lie within one step of the ground-truth configuration. The step size is parameter-specific, which is 1 for $L$, 128 for $H$, 4 for $A$, and 1024 for $I$.

We implement a minimal autoregressive inference loop using the standardized Meta Llama implementation provided by the HuggingFace \texttt{Transformers} library~\cite{wolf-etal-2020-transformers}. To ensure accurate per-token timing during data collection in both offline and online phases, we insert exactly two CUDA synchronization instructions, one before starting the timer and one before stopping it, thereby isolating the generation time of each token without interference from preceding or subsequent token generations. This synchronization overhead is minimal and does not disrupt GPU execution during single-token inference. We use Python’s standard \texttt{time} library to log timestamps at the necessary points in the inference loop. For per-subcomponent timing measurements in the offline phase, we introduce conditional checks on control parameters that allow enabling or bypassing individual transformer sub-components. This modification does not affect the timing behavior of the model as a whole but effectively removes the contribution of a specific sub-component and the overhead imposed by it from the overall runtime. We run the runtime prediction and architectural classification evaluations on a 40-core Intel(R) Xeon(R) Silver 4416 CPU.

The following are the specific setups used for different experiments in this section:

\niparagraph{\textbf{Eager Attention Experiments}.}
In addition to the test dataset generated from the default Llama~3.2~3B target model, we generate an additional test dataset using the 1B model to evaluate in-domain generalization of the predictor model. When generating test data from the 1B model, we restrict the configurations to those not used in the training dataset. The offline collection of training and testing runtime data yields measurements for 81 unique ($H$, $L$) configurations in the training set, 110 configurations in the 1B-model test set, and 130 configurations in the 3B-model test set.

Both the reference and target models are executed on a single NVIDIA GeForce RTX~2080~Ti GPU (11~GB VRAM) with CUDA~12.4 and cuBLAS enabled, minimizing CPU–GPU transfer and PCIe overheads. We use 24 identical GPUs distributed across three servers, each running the same model and experiment on a distinct data partition. To ensure that no attention optimization is applied during these experiments, we explicitly force the model to use the eager attention implementation, with KV-cache disabled.

\niparagraph{\textbf{Open Model Experiments}.}
To evaluate cross-family generalization, we generate new test datasets using three different open models: the 1.5B model from the Qwen2.5 family~\cite{qwen2.5}, the Phi3.5-mini~3.8B model~\cite{abdin2024phi3technicalreporthighly}, and the 2B model from the Gemma2 family~\cite{gemmateam2024gemma2improvingopen}. The resulting test datasets consist of 44, 40, and 40 distinct ($H$, $L$) configurations, respectively. We run these experiments on the same RTX~2080~Ti setup as the eager attention experiments.

\niparagraph{\textbf{FlashAttention Experiments}.}
For experiments involving FlashAttention2, which requires Ampere or newer GPUs, we use an NVIDIA A10 GPU (24~GB VRAM) with CUDA~12.6 to generate training and test datasets consisting of 81 and 130 configurations, respectively. During inference, we force the \texttt{transformers} library to use PyTorch's built-in scaled dot-product attention (SDPA) FlashAttention2~\cite{dao2023flashattention2fasterattentionbetter} implementation.

\niparagraph{\textbf{KV+FlashAttention Experiments}.}
For experiments with KV-cache–enabled inference, we use the same inference setup as in the FlashAttention experiments, with KV caching enabled. These experiments are run on an NVIDIA B200 GPU (180~GB HBM3E VRAM) from the Blackwell series, with CUDA~13.0, using extended prompt lengths for both the Llama~1B and 3B models. We construct the training and test datasets following the same methodology, consisting of 159 and 148 configurations, respectively.

\niparagraph{\textbf{Remote Inference API Experiment}.}
To evaluate the accuracy of our runtime prediction framework and the feasibility of our architectural leakage attack in real-world deployments, we collect per-token timing traces from the Weights \& Biases Llama~3.1~8B Instruct streaming API~\cite{wandb-llama31-8b-instruct-inference}. In our evaluations, we treat the API as a fully unknown black-box model without access to its internal architectural specifications. For reporting the results, we subsequently use the documented model configuration to validate the inferred parameters and runtime prediction accuracy. Although the target model belongs to the same family used for training, its architectural dimensions lie outside the range of parameters seen by the predictor during training.  {In this experiment, the target model is hosted on a remote machine to which we do not have direct access and is served using an unknown, unmodified inference framework. Unlike our local experiments, we do not instrument the serving stack for timing collection. The per-token timing collection client follows the same setup described in \S\ref{sec:attack1-remote-setup}.}

We select this API from a range of providers to minimize the impact of unmodeled optimizations and to ensure that the model is served in its original FP16 precision.

\subsection{Results}
\label{sec:attack2-results}

\begin{figure}[t]
  \centering
  \includegraphics[width=\columnwidth]{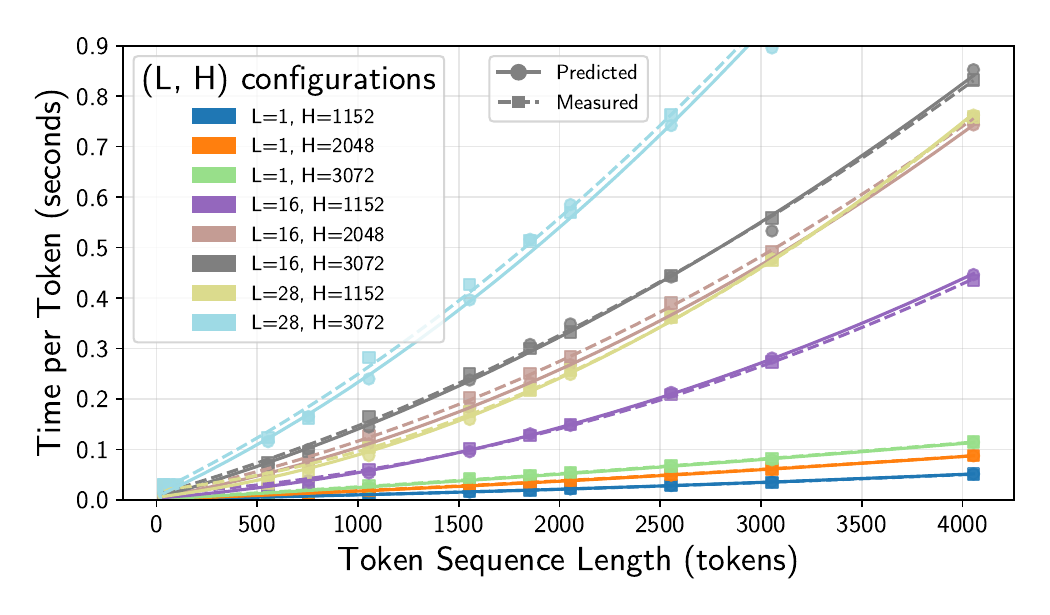}
  \caption{\textbf{Predicted token-generation times vs. ground-truth times. }\textit{Our model can accurately predict the per-token generation time for various configurations.}}
  \label{fig:attack2-results-predict}
\end{figure}

\begin{table}[t]
\centering
\caption{Accuracy of the runtime predictor model, measured as normalized root mean squared error (NRMSE).}
\label{tab:nrmse1}
\setlength{\tabcolsep}{4pt}
\renewcommand{\arraystretch}{1}
\resizebox{\columnwidth}{!}{%
\begin{tabular}{@{}l@{\hspace{20pt}}*{8}{r}@{}}
\toprule
\multirow{2}{*}{Step}
& \multicolumn{2}{c}{Eager (Naive)}
& \multicolumn{2}{c}{Eager (Corrected)}
& \multicolumn{2}{c}{Flash2}
& \multicolumn{2}{c}{KV-Flash2} \\
\cmidrule(lr){2-3}
\cmidrule(lr){4-5}
\cmidrule(lr){6-7}
\cmidrule(l){8-9}
& Train & Test & Train & Test & Train & Test & Train & Test \\
\midrule

Decode
& 0.106 & 0.411
& 0.106 & 0.119
& 0.258 & 0.209
& 0.218 & 0.252 \\

Prefill
& - & -
& - & -
& - & -
& 0.168 & 0.188 \\

% \midrule

% L
% & 100.00 & 69.23
% & 3 & 4
% & 80.56 & 97.69
% & 100.00 & 83.78 \\

% H
% & 100.00 & 14.62
% & 3 & 4
% & 99.07 & 100.00
% & 89.94 & 70.95 \\

% (H, L)
% & 100.00 & 27.69
% & 91.36 & 65.38
% & 73.15 & 54.62
% & 72.96 & 45.27 \\

\bottomrule
\end{tabular}%
}
\end{table}

\niparagraph{\textbf{Accuracy of Token Generation Time Prediction}.} We first assess the ability of our models in predicting the token generation time of models with different configurations.

Figure~\ref{fig:attack2-results-predict} illustrates the predicted runtimes along with the corresponding ground-truth measurements for several test configurations using eager attention.
The results highlight the quadratic relationship between the runtime and the sequence length, and show that our simple regression-based model accurately captures this behavior.
Even for seemingly similar timing patterns, such as (L=28, H=1152) and (L=16, H=2048), we observe distinct scaling behavior. The former configuration (more layers) starts with lower generation latency but exhibits higher growth rate, eventually surpassing the latter configuration (larger hidden dimension).
This further supports our claim that, given a sufficiently large range of input sequences and precise timings, it is possible to uniquely distinguish between different architectural parameters.
We present similar example configurations demonstrating these patterns for \texttt{Flash2} and \texttt{KV-Flash2} models in Appendix \S\ref{sec:appendix-flash} and \S\ref{sec:appendix-flash-kv}.

% \begin{table*}[t]
% \centering
% \caption{NRMSE and Top-N comparison across different predictors.}
% \label{tab:attack2_kernel_results}
% \begin{tabular}{l|cc|cc|cc|cc}
% \toprule
%  & \multicolumn{2}{c|}{Naive} 
%  & \multicolumn{2}{c|}{Corrected} 
%  & \multicolumn{2}{c|}{Flash2} 
%  & \multicolumn{2}{c}{KV-Flash2} \\
% \cline{2-9}
%  & Train & Test & Train & Test & Train & Test & Train & Test \\
% \midrule
% NRMSE   & 0.106 & 0.411 & 0.106 & 0.119 & 0.258 & 0.209 & 0.168/0.218 & 0.188/0.252 \\
% \midrule
% Top-N (H fixed)   & 100.00/100.00 & 90.77/69.23 & 3 & 4 & 90.74/80.56 & 100.00/97.69 & 100.00/100.00 & 95.27/83.78 \\
% Top-N (L fixed)   & 100.00/100.00 & 56.15/14.62 & 3 & 4 & 99.07/99.07 & 100.00/100.00 & 100.00/89.94 & 98.65/70.95 \\
% Top-N (H, L)      & 100.00/100.00 & 33.08/27.69 & 100.00/91.36 & 83.85/65.38 & 84.26/73.15 & 75.38/54.62 & 86.79/72.96 & 64.86/45.27 \\
% \bottomrule
% \end{tabular}
% \end{table*}

\begin{table}[t]
\centering
\footnotesize
\caption{Top-5 accuracy (\%) of parameter leakage attack.}
\label{tab:top5_attack2_kernel_results}
\setlength{\tabcolsep}{4pt}
\renewcommand{\arraystretch}{1}
% \resizebox{\columnwidth}{!}{%
\begin{tabular}{l@{\hspace{20pt}}*{6}{r}}
\toprule
\multirow{3}{*}{\shortstack[l]{Target\\Parameter}}
% & \multicolumn{2}{c}{Eager (Naive)}
& \multicolumn{2}{c}{Eager (Corrected)}
& \multicolumn{2}{c}{Flash2}
& \multicolumn{2}{c}{KV-Flash2} \\
\cmidrule(lr){2-3}
\cmidrule(lr){4-5}
\cmidrule(lr){6-7}
% \cmidrule(l){8-9}
% & Train & Test 
& Train & Test & Train & Test & Train & Test \\
\midrule

L
% & 100.00 & 69.23
& 95.06 & 86.15
& 80.56 & 97.69
& 100.00 & 83.78 \\

H
% & 100.00 & 14.62
& 100.00 & 71.54
& 99.07 & 100.00
& 89.94 & 70.95 \\

(H, L)
% & 100.00 & 27.69
& 91.36 & 65.38
& 73.15 & 54.62
& 72.96 & 45.27 \\

\bottomrule
\end{tabular}%
% }
\end{table}

Table~\ref{tab:nrmse1} shows the normalized root mean squared error (NRMSE) of token-generating time prediction for train and test datasets for various attention implementations.

To evaluate the effectiveness of our correction for non-linearities introduced by CuBLAS kernel selection in eager attention (\S\ref{sec:online-runtime-correction}), we report results for both \texttt{Eager(Naive)} and \texttt{Eager(Corrected)} variants.
Both variants achieve low error on the train dataset generated by Llama~1B model. 
However, \texttt{Eager(Naive)}'s performance plummets on the test dataset generated by the 3B model, while the \texttt{Eager(Corrected)} variant maintains its accuracy, showing the effectiveness of our non-linearity correction in generalizing to unseen architectures.

The table also shows that our \texttt{Flash2} and \texttt{KV-Flash2} can also predict the token generation time of unseen architectures, albeit with lower accuracy than the simple \texttt{Eager} predictor.
This reduced accuracy stems from the highly fused and complex nature of the FlashAttention2 kernel, whose runtime behavior depends on multiple interacting factors beyond the dominant memory access term captured in our model. While we incorporate the primary scaling effect introduced by FlashAttention, additional lower-order factors and implementation-specific details are not explicitly modeled. Furthermore, discrepancies between the theoretical specification and practical implementations introduce small variations in runtime that are not fully captured by our scaling terms.

For inference with KV-cache enabled, decoding processes one token per step, resulting in timing signals that are less informative due to limited variation across sequence lengths.
In contrast, the prefill stage, where the KV-cache is constructed, is heavily influenced by context length and architectural parameters, and thus provides a more useful signal.
Therefore, for this predictor we also model prefill timings as well. 
The results in Table~\ref{tab:nrmse1} show that our model can accurately capture the timing of KV-cache prefill, achieving an NRMSE of 0.18 on the test dataset.

\begin{table}[t]
\centering
\caption{NRMSE across different test models.}
\label{tab:attack2_model_nrmse}
\setlength{\tabcolsep}{4pt}
\renewcommand{\arraystretch}{0.95}
\resizebox{\columnwidth}{!}{%
\begin{tabular}{@{}l*{5}{r}@{}}
\toprule
Model
& Llama3.2 1B
& Llama3.2 3B
& Qwen2.5 1.5B
& Phi3.5-mini 3.8B
& Gemma2 2B \\
\midrule

NRMSE
& 0.221
& 0.119
& 0.159
& 0.183
& 0.186 \\

\bottomrule
\end{tabular}%
}
\end{table}

\niparagraph{\textbf{Accuracy of Model Parameter Leakage Attack}.}
We next evaluate how our models can be used to infer the architectural dimensions based on the observed timing traces.
As discussed in \S\ref{sec:attack2-online-phase}, we use our runtime predictor model to build a classifier that ranks the candidate architectural configurations for a given observed timing trace, based on their similarity to the trace generated by the predictor.

We classify each timing trace in the datasets to one of the classes in a large configuration space (discussed in \S\ref{sec:attack2-exp-setup}), and report the top-5 accuracy of this classification.
We repeat this for different set of target parameters that we aim to leak.
Table~\ref{tab:top5_attack2_kernel_results} presents the results of these experiments.

On the unseen test datasets, the attack can achieve a high top-5 accuracy ($>70\%$) across all predictors when inferring only one unknown architectural parameter. 
When both H and L are unknown, the task becomes significantly more challenging, as the label space is much larger.
In addition, most asymptotic terms in our runtime model depend on the product $H\cdot L$, making it difficult for the classifier to separate their individual impact. 
The lower accuracy of our runtime predictor for \texttt{Flash2} and \texttt{KV-Flash2}, compared to the \texttt{Eager} variant as discussed above, directly impacts the classification accuracy.
Therefore, the classification accuracy of these models is generally lower than the \texttt{Eager} predictor.

% \begin{table*}[t]
% \centering
% \caption{NRMSE and Top-N comparison across different models.}
% \label{tab:attack2_model_results}
% \begin{tabular}{l|c|c|c|c|c}
% \toprule
%  & Llama3.2 1B
%  & Llama3.2 3B
%  & Qwen2.5 1.5B
%  & Phi3.5-mini 3.8B
%  & Gemma2 2B \\
% \midrule
% NRMSE   & 0.221 & 0.119 & 0.159 & 0.183 & 0.186 \\
% \midrule
% Top-N (H fixed)   & 100.00/100.00 & 99.23/86.15 & 97.73/77.27 & 100.00/97.50 & 100.00/100.00  \\
% Top-N (L fixed)   & 100.00/93.64 & 96.92/71.54 & 100.00/100.00 & 100.00/97.50 & 100.00/77.50  \\
% Top-N (H, L)      & 94.55/90.91 & 83.85/65.38 & 77.27/50.00 & 90.00/80.00 & 72.50/47.50   \\
% \bottomrule
% \end{tabular}
% \end{table*}

\begin{table}[t]
\centering
\caption{Top-5 accuracy (\%) across different test models.}
\label{tab:attack2_model_top5}
\setlength{\tabcolsep}{4pt}
\resizebox{\columnwidth}{!}{%
\begin{tabular}{@{}l@{\hspace{20pt}}*{5}{r}@{}}
\toprule
Target
& \multicolumn{1}{c}{Llama3.2 1B}
& \multicolumn{1}{c}{Llama3.2 3B}
& \multicolumn{1}{c}{Qwen2.5 1.5B}
& \multicolumn{1}{c}{Phi3.5-mini 3.8B}
& \multicolumn{1}{r}{Gemma2 2B} \\
\midrule

L
& 100.00
& 86.15
& 77.27
& 97.50
& 100.00 \\

H
& 93.64
& 71.54
& 100.00
& 97.50
& 77.50 \\

(H, L)
& 90.91
& 65.38
& 50.00
& 80.00
& 47.50 \\

\bottomrule
\end{tabular}%
}
\end{table}

\niparagraph{\textbf{Cross-Family Generalization}.}
We next evaluate whether a predictor trained on one model family generalizes to decoder-only architectures from different model families. We reuse the \texttt{Eager(Corrected)} predictor trained on timing data from the Llama~1B model and evaluate its performance on test datasets generated from our three representative open models.
Table~\ref{tab:attack2_model_nrmse} shows the runtime prediction accuracy for these models, and Table~\ref{tab:attack2_model_top5} shows the  top-5 success rate of the architectural configuration leakage attack.
The high success rate for different models indicates that our linear regression-based prediction framework does not overfit to a specific model family and generalizes well across architectures. Additional details on prediction accuracy for the open models are presented in Appendix \S\ref{sec:appendix-cross}.

\begin{figure}[t]
  \centering
  \includegraphics[width=\columnwidth]{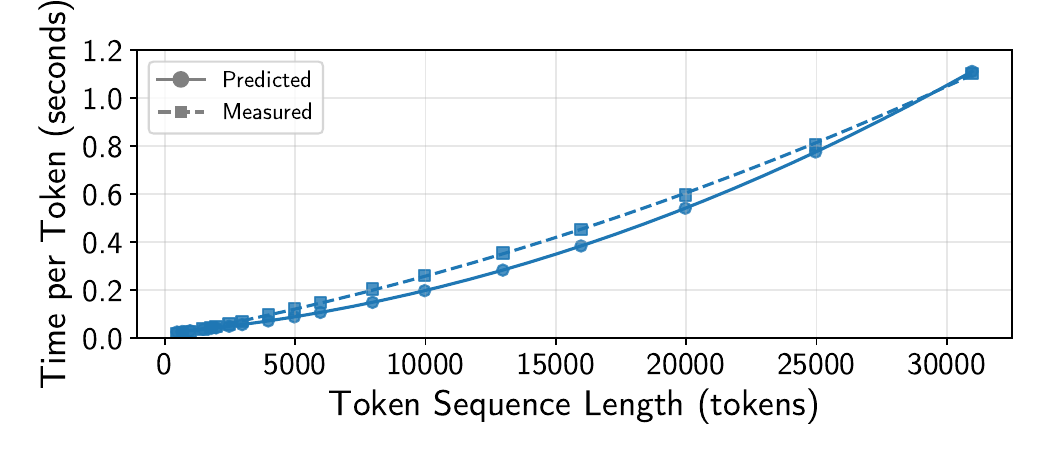}
  \caption{\textbf{Predicted token-generation times vs. ground-truth prefill stage times for W\&B's Llama 3.1 8B Instruct remote API. }}
  \label{fig:attack2-remote-predict}
\end{figure}

\niparagraph{\textbf{Leaking Architectural Parameters from a Remote Inference API}.}
Next, we evaluate our attack in leaking architectural parameters of an unknown black-box model accessible only through a remote inference service API.
First, we use our \texttt{KV-Flash2} predictor to estimate the expected timing behavior of a model with the same architecture as our remote Llama3.1~8B target (on Weights\&Biases).
As shown in Figure~\ref{fig:attack2-remote-predict}, the predicted timings closely match the observed measurements from the remote API generation times.
Next, we perform the architectural parameter recovery experiment over an extended $[H, L]$ search grid to avoid bias toward larger configurations. When assuming   $L$ is known and inferring $H$, the correct value is recovered as the top-ranked candidate.
When fixing $L$ and inferring $H$, the attack ranks the correct configuration within the top-5 (ranked 4th). 
When searching over both dimensions, a near match $(H=4096, L=34)$ is ranked 8/2068, and the true configuration $(H=4096, L=32)$ is ranked 28/2068. 
These results demonstrate that our approach can leak architectural parameters in realistic API settings.
% }

%-------------------------------------------------------------------------------
%%  Discussion
%-------------------------------------------------------------------------------
%-------------------------------------------------------------------------------
% \section{Discussion}
% %-------------------------------------------------------------------------------

\section{Mitigation}
Mitigating timing leakage without compromising QoS is challenging. Below, we discuss potential mitigation strategies.

\niparagraph{\textbf{Constant Per-Token Time}.} 
Our attacks exploit per-token timing information that depends on both model and deployment characteristics.
In particular, for the model leakage attack, due to the autoregressive nature of transformers, later tokens typically take longer to generate than earlier ones, and the rate of this increase reveals information about key model dimensions.
To defend against such attacks, the ideal approach would be to eliminate the dependency of generation time on model parameters. However, achieving this in practice would require making all token generations as slow as the worst-case scenario, potentially increasing latency by several times.
This mitigation strategy would therefore defeat the purpose of inference optimizations entirely.

\niparagraph{\textbf{Buffering}.} 
An alternative defense is to reduce timing granularity by buffering token outputs instead of streaming them immediately. For example, the provider could aggregate a fixed number of tokens N and transmit them at constant intervals T, thereby flattening timing variations. Yet, this approach may be impractical for large models or latency-sensitive applications, since delaying token transmission can degrade Quality of Service (QoS) and user experience.

%-------------------------------------------------------------------------------
%%  Related Work
%-------------------------------------------------------------------------------
%-------------------------------------------------------------------------------
\section{Related Work}
\label{sec:related}
%-------------------------------------------------------------------------------

\niparagraph{\textbf{Side-Channel Attacks on Large Language Models}.}
Prior works have demonstrated attacks that exploit various side channels, including timing~\cite{duddu2019stealingneuralnetworkstiming}, cache and memory access patterns~\cite{hua2018reverse,CacheTelepathy}, power consumption~\cite{gao2024deeptheft}, electromagnetic emissions~\cite{batina2019csi}, and GPU-level side channels~\cite{wei2020leaky}, to extract model weights, output labels, or hardware properties from classical deep neural networks.

Side-channel attacks have also emerged for modern LLMs. Recent works exploit timing channels stemming from inference-time optimizations~\cite{carlini2024remote}, caching and memory behaviors~\cite{erlyTime2025,zheng2024inputsnatchstealinginputllm,gu2025auditingpromptcachinglanguage}, network-level artifacts such as token-count or packet-length side channels~\cite{mcdonald2025whisper,zhang2024timetelltimingchannels,wei2025speculationspillssecretschannels}, and cache-based attacks~\cite{KnowSaidCache}. These efforts primarily focus on fingerprinting models, leaking user–LLM interaction details under specific conditions, or, in some cases, recovering limited architectural or deployment information, such as through GPU performance counters~\cite{xu2025wave} or long-range EM leakage~\cite{xiao2026peering}.

There also exist model-stealing attacks that don't rely on side channels. Carlini et al.~\cite{carliniStealProd} recover the hidden dimension size \(H\) and projection layer \(W_{proj}\) of production models with near-exact precision by exploiting the returned logit bias from the API. Compared to our architecture-leakage attack in \S\ref{sec:attack2-leakarch}, their attack relies on extra information returned by the API and is more costly, since all interaction must occur online.

\niparagraph{\textbf{Modeling of Large Language Models}.}
% \kazem{this section can go if we need space}
Several previous works have developed performance-modeling tools and analytical frameworks for understanding or predicting LLM behavior, including trace-based simulation frameworks~\cite{tracesim}, analytical models for distributed training and inference~\cite{distPerfModel}, forecasting frameworks for model performance on unseen GPUs~\cite{lee2024neusight}, and hardware-agnostic analytical modeling of LLM inference~\cite{patwari2025forecastingllminferenceperformance}. Accurately modeling GPUs is crucial for insightful performance models for LLMs. Accel-Sim~\cite{accelsim} provides a detailed GPU simulation environment, while work such as~\cite{gpu_component} offers a component-wise perspective on GPU execution costs. Amali~\cite{amalimodel} provides analytical modeling for inference workloads on modern GPUs. However, most of these approaches adopt a system-centric view (e.g., focusing on memory transactions, kernel launches) and lack the theoretical parameters governing LLM scaling laws. Our work bridges this gap by combining theoretical scaling terms with empirical measurements.

%-------------------------------------------------------------------------------
%%  Conclusion
%-------------------------------------------------------------------------------
%-------------------------------------------------------------------------------
\section{Conclusion}
%-------------------------------------------------------------------------------
This paper presents LeakyLMs, the first systematic study demonstrating that fine-grained per-token generation timings from production large language models can leak sensitive information about both model architecture and deployment strategies.
This poses a significant threat, as such information provides a competitive advantage to leading organizations in the ongoing AI race.
Defending against such timing side channels without degrading latency or user experience remains an open and difficult challenge.
This work represents an initial step toward understanding and mitigating timing-based model leaks in large language model deployments.

%-------------------------------------------------------------------------------
%%  Acknowledgement
%-------------------------------------------------------------------------------
\begin{acks}
The authors would like to gratefully acknowledge Longview Philanthropy for their  support of our ongoing research in this area. 
\end{acks}

%-------------------------------------------------------------------------------

\cleardoublepage
\appendix

\bibliographystyle{ACM-Reference-Format}
\bibliography{body/references}

%%
%% Appendices
\appendix
\section{Additional Details on Leaking Inference Optimizations Attack}

\subsection{Network Effect on Time Measurements}
\label{appendix-spec-net}

Prior to measurement, we profile the network to establish baseline latency and noise. For API access, we use each provider’s official Python package when available, otherwise we use Python's \texttt{requests} library.  
Our measurements for the two remote models in Table~\ref{table:attack1_exp_setup} show that the average round-trip time (RTT) for Gemini and Cohere is 26.65~ms and 5.75~ms, respectively, with RTT standard deviations of 0.07~ms and 0.08~ms. Note that a constant RTT between our client and the APIs only produces a fixed offset in the timing traces. It does not alter the relative per-token timing pattern. 
The principal network effect that can confound our measurements is RTT variability. If the standard deviation of arrival times is on the same order of magnitude as the per-token generation time, the introduced noise can obscure the LLM’s true timing behavior. The measured RTT standard deviations for our tested remote models are below 5\% of the minimum per-token generation time.

\subsection{Remote Black-Box Model Results}
\label{appendix-spec-A}

Table~\ref{table:attack1_results_detection} summarizes all the tested models for inference optimization and the detection outcomes.

\begin{table}[t]
\centering
\caption{Results of speculative decoding detection across different LLM models.}
\label{table:attack1_results_detection}
\begin{tabularx}{\columnwidth}{@{} l *{1}{>{\centering\arraybackslash}X} @{}}
  \toprule
  \textbf{LLM Model} & \textbf{Uses Speculative Decoding?} \\ 
  \midrule
    LLaMA2(Guanaco) 13B + TinyLlama 1.1B & \cmark \\
    Gemini 2.5 Pro & \_ \\
    Gemini Flash 2.5 & \cmark \\
    Gemini Flash 2.5 Lite & \cmark \\
    Gemini Flash 2.0 & \_ \\
    Gemini Flash 2.0 Lite & \_ \\
    Gemini Flash 1.5 & \cmark \\
    OpenAI gpt-4o & \_ \\
    OpenAI gpt-4-0613 & \_ \\
    OpenAI gpt-3.5-turbo & \_ \\
    Mistral large-2411 & \_ \\
    Cohere command-r-plus-08-2024 & \_ \\
    Perplexity sonar & \_ \\
    Cerebras qwen-3-235b-a22b-instruct-2507 & \_ \\
    Groq llama3-8b  & \_ \\
    
  \bottomrule
\end{tabularx}
\end{table}

\begin{figure}[t]
  \centering
  \includegraphics[width=\columnwidth]{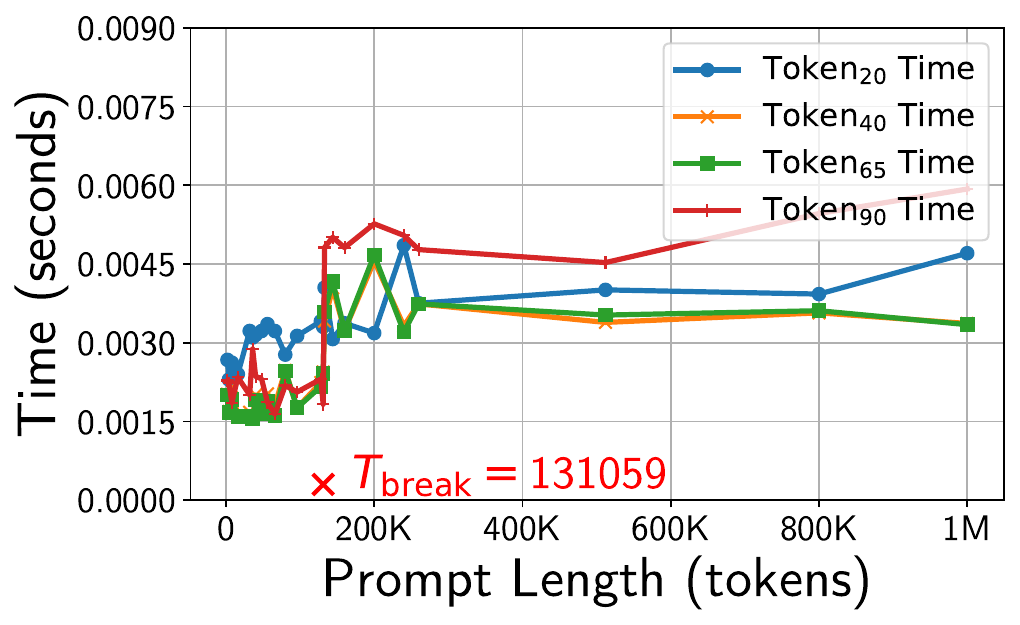}
  \caption{\textbf{Per-token generation time of Gemini Flash-2.5-Lite exhibiting speculative decoding behavior.}}
  \label{fig:attack1-results-extra-gemini25-lite}
\end{figure}

Figure~\ref{fig:attack1-results-extra-gemini25-lite} depicts the successful detection of the speculative decoding signature timing jump in Gemini Flash-2.5-Lite. The gap between the two extremes of the timing jump is slightly smaller than in other successful cases. This may be due to the smaller difference in size between the main and draft models.

\begin{figure}[t]
  \centering

  \begin{subfigure}[t]{0.49\columnwidth}
      \centering
      \includegraphics[width=\textwidth]{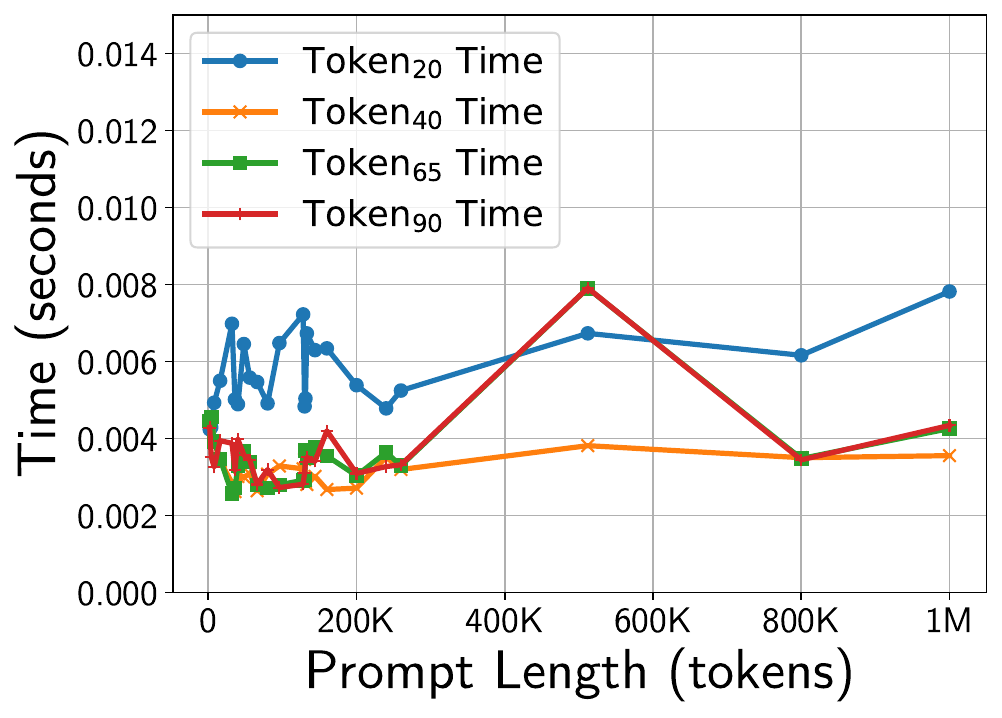}
      \caption{Gemini Flash-2.0-Lite}
      \label{fig:attack1-results-remote-flash20-lite}
  \end{subfigure}
  \hfill
  \begin{subfigure}[t]{0.49\columnwidth}
      \centering
      \includegraphics[width=\textwidth]{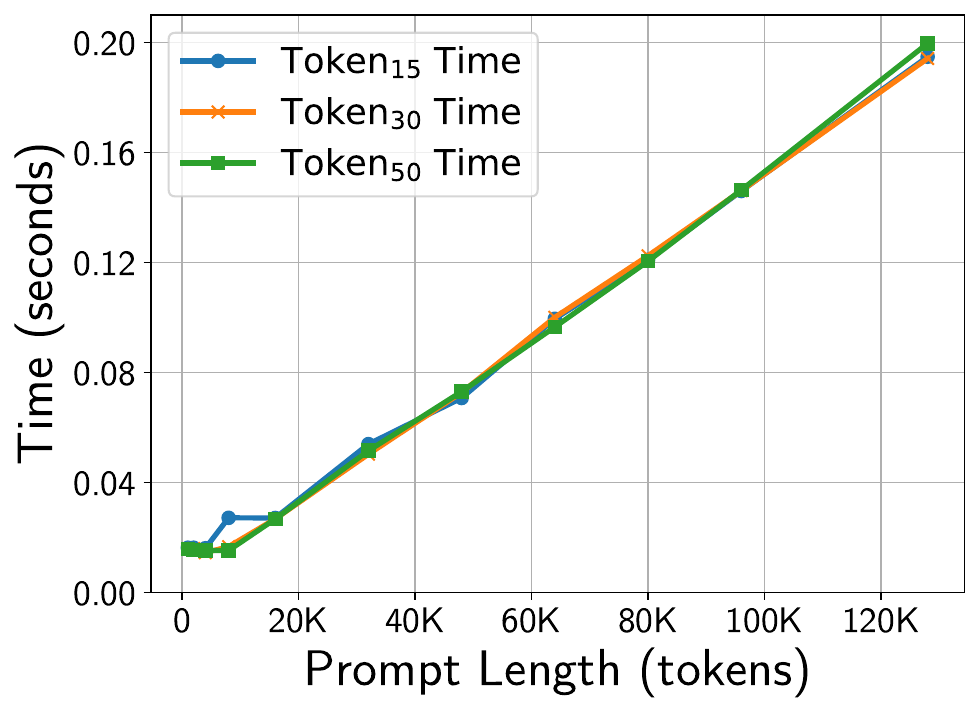}
      \caption{Cohere Command R+}
      \label{fig:attack1-results-remote-cohere-rplus}
  \end{subfigure}

  \caption{\textbf{Per-token generation time of models where speculative decoding was not detected.}}
  \label{fig:attack1-results-remote-fail}
\end{figure}

Figure~\ref{fig:attack1-results-remote-fail} shows examples of timing responses from APIs where no speculative decoding signature was detected. As shown in Figure \ref{fig:attack1-results-remote-cohere-rplus}, the Cohere Command R+ model exhibits a non-constant yet stable per-token timing pattern, likely reflecting natural variations due to KV-cache utilization rather than speculative decoding. In contrast, other models with no detected speculative behavior, such as Gemini Flash-2.0-Lite (Figure \ref{fig:attack1-results-remote-flash20-lite}), show nearly constant and consistent timing across input lengths. These observations confirm that our method depends on the consistency of timing rather than model-specific scaling laws affected by other optimizations, thereby amplifying its robustness and reducing the likelihood of false positives.

\section{Additional Details on Leaking Model Architecture Attack}

\subsection{Details of Scaling Runtime for Unseen Dimensions}
\label{sec:app-online-runtime-correction}

\begin{figure}
    \centering
    \includegraphics[width=1\linewidth]{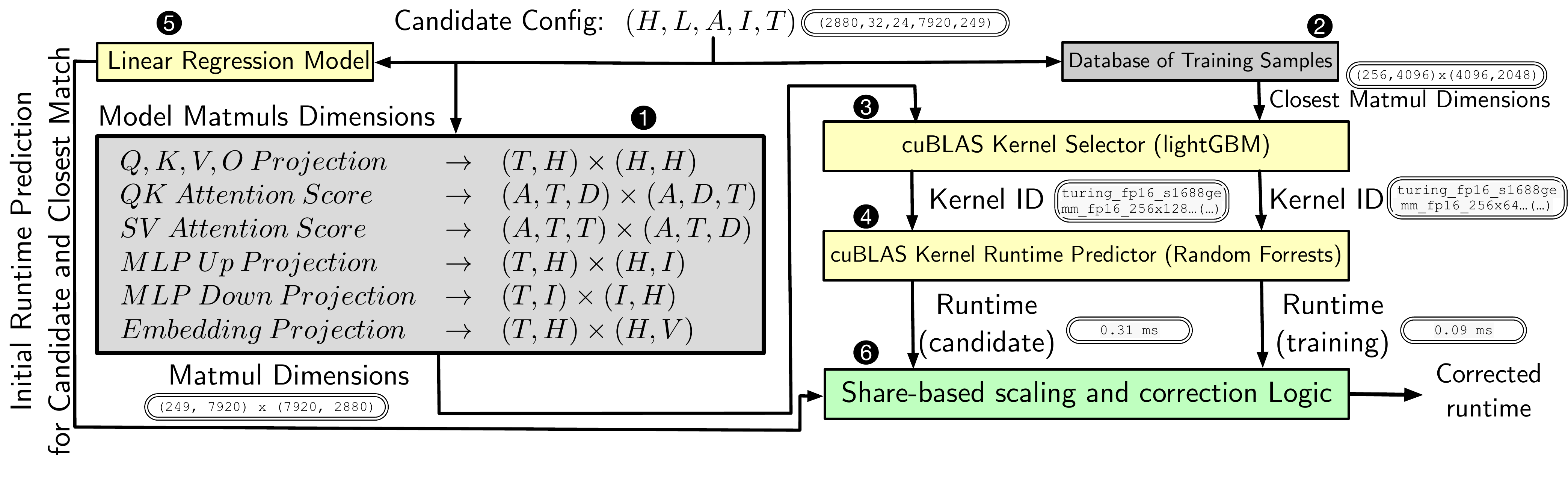}
    \caption{\textbf{Non-linearity correction model. }\textit{We use two predictors to predict the runtime of an unseen matmul dimension and correct the output of our model.}}
    \label{fig:attack2-online-matmul_correction}
\end{figure}

Figure~\ref{fig:attack2-online-matmul_correction} shows our method for correcting the nonlinearities that arise from varying CUDA kernel selections described in~\S\ref{sec:attack2-overview}.
%
% We use the same illustration below to describe the correction steps for a candidate architectural configuration and input length pair that the predictor has not seen during training.
Assume a candidate architectural configuration and input length (\(H\), \(L\), \(A\), \(I\), \(T\)). When applying the correction pipeline to the original runtime predictor, we take the following steps:

\begin{CompactItemize}
    \item In step \ding{192}, the candidate configuration is expanded into the full set of matmul operand shapes required to be executed to generate one output token for a prompt of length $T$ and an LLM with the dimensions \((H, L, A, I)\).

    \item In step  \ding{193}, we query the original training dataset and retrieve the closest configuration–prompt pair to the candidate. We also enumerate the matmul operand shapes for that closest sample.

    \item In step  \ding{194}, for each matmul in both the candidate and the closest configurations, we feed the operand dimensions into the LightGBM kernel classifier (trained in the offline phase) to predict which cuBLAS kernel is likely to be selected.
    % For example, the MLP down-projection matmul of the candidate with dimensions (249,7920)×(7920,2880) is predicted to use a \texttt{256x128 gemm} kernel, whereas the corresponding matmul in the closest sample (256,4096)×(4096,2048) uses \texttt{256x64 gemm}.
    
    \item In step  \ding{195}, for each matmul and its predicted kernel in \ding{194}, we use the kernel-specific random-forest regressor to predict a heuristic runtime. Denote these as  $heuristic_{cand}$ and $heuristic_{close}$ for candidate and closest matmuls, respectively.    

    \item In step  \ding{196}, we run the original linear regression predictor on both the candidate and the closest configuration to obtain initial predicted runtimes \(time_{cand}\) and \(time_{close}\), respectively.

    \item In step \ding{197}, for every matmul that requires correction, we compute a scaling factor \( f =\frac{heuristic_{cand}}{heuristic_{close}}\). We then scale the closest sample’s initial linear-predicted runtime \(time_{close}\) for that matmul by \(f\). To preserve consistency of this component-wise correction approach, we apply only the portion of this scaled time that is contributed by the target matmul, i.e., \( f.time_{close}.share \), where share is approximated by the matmul’s OPs divided by the total OPs of the block. If a matmul is marked as not needing correction, we instead use its share from the candidate’s initial linear prediction \(time_{cand}.share\). After applying this procedure to all matmuls and summing their adjusted contributions, we obtain the final corrected runtime for the component.
    
\end{CompactItemize}

\subsection{Detailed Theoretical Terms for All Operations}
\label{sec:appendix-A-1}

Following the theoretical analysis of runtime scaling terms per operation described in \S\ref{sec:attack2-method-features}, we present the complete list of theoretical scaling terms extracted for each operation. These runtime scaling terms apply to a single decoder only.

\noindent
\textbf{Q/K/V/O projection}:

\begin{equation}
\begin{split}
time &= \alpha_1 T H^2 d + \alpha_2 H^2bc + \alpha_3 THbc
\end{split}
\end{equation}

\noindent
\textbf{S = QK and Att = SV attention multiplications}:

\begin{equation}
\begin{split}
time &= \alpha_1 T^2 H d + \alpha_2 THbc + \alpha_3 T^2Abc
\end{split}
\end{equation}

\noindent
\textbf{attention softmax}:

\begin{equation}
\begin{split}
time &= \alpha_1 T^2 A d + \alpha_2 T^2Abc
\end{split}
\end{equation}

\noindent
\textbf{MLP up/down projections}:

\begin{equation}
\begin{split}
time &= \alpha_1 T H I d + \alpha_2 THbc + \alpha_3 HIbc + \alpha_4 TIbc
\end{split}
\end{equation}

\subsection{Detailed Theoretical Terms for All Subcomponents}
\label{sec:appendix-A-2}

Below, we compose the scaling terms for the subcomponents from their underlying operations, accounting for the contributions of all decoder layers.

\noindent
\textbf{multi-head (eager) attention subcomponent}:

\begin{equation}
\begin{split}
time_{Att} &= \alpha_1 T H^2 L d + \alpha_2 T^2 H L d \\
           &\quad + \alpha_3 T^2 A L d + \alpha_4 H^2 L b c \\
           &\quad + \alpha_5 T^2 A L b c + \alpha_6 T H L b c
\end{split}
\end{equation}

\begin{equation}
time_{Att} \sim T^2, H^2, A, L
\end{equation}

\noindent
\textbf{MLP subcomponent}:

\begin{equation}
\begin{split}
time_{MLP} &= \alpha_1 THILd + \alpha_2 HILbc \\
           &\quad + \alpha_3 TILbc + \alpha_4 THLbc
\end{split}
\end{equation}

\begin{equation}
time_{MLP} \sim T, H^2, I, L
\end{equation}

\noindent
\textbf{add and normalization subcomponent}:

\begin{equation}
\begin{split}
time_{Norm} &= \alpha_1 THLd + \alpha_2 HLbc \\
           &\quad + \alpha_3 THLbc
\end{split}
\end{equation}

\begin{equation}
time_{Norm} \sim T, H, L
\end{equation}

\noindent
\textbf{embedding/projection (naive) subcomponent}:

\begin{equation}
\begin{split}
time_{EmbedProj} &= \alpha_1 THVd + \alpha_2 HVbc \\
           &\quad + \alpha_3 THbc
\end{split}
\end{equation}

\begin{equation}
time_{EmbedProj} \sim T, H, V
\end{equation}

The term $THVd$ in EmbedProj can be replaced with $HVd$ depending on whether the implementation projects the entire sequence or only the newly generated token in each round.

\subsection{Cross-Family Generalization}
\label{sec:appendix-cross}

\begin{figure}[t]
  \centering
  \includegraphics[width=\columnwidth]{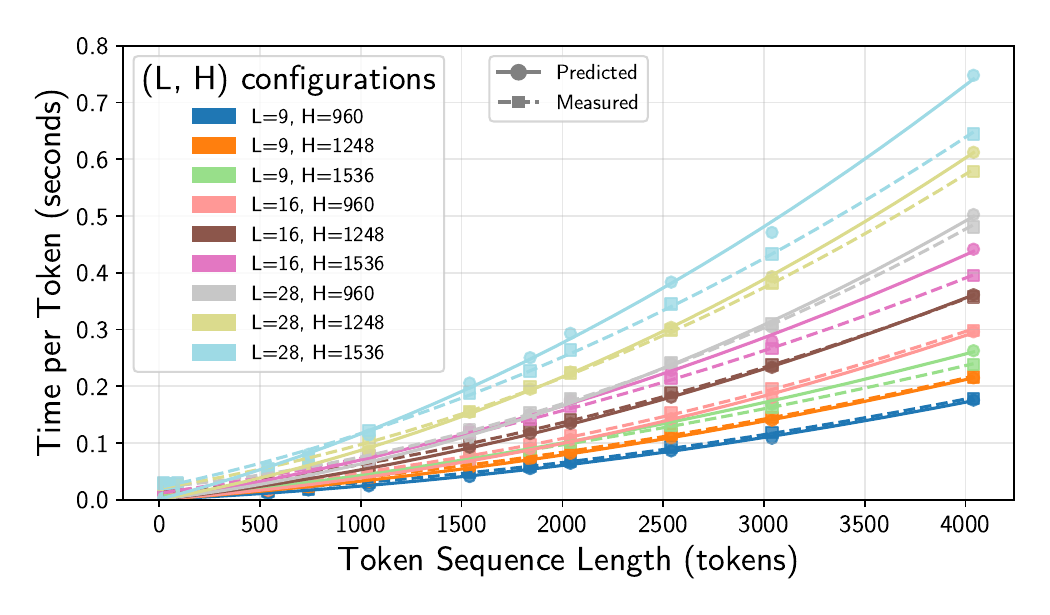}
  \caption{\textbf{Predicted token-generation times (generated by a predictor trained on Llama~1B) vs. ground-truth times (obtained from inference on the Qwen2.5~1.5B model).}}
  \label{fig:qwen-pred-res}
\end{figure}

\begin{figure}[t]
  \centering
  \includegraphics[width=\columnwidth]{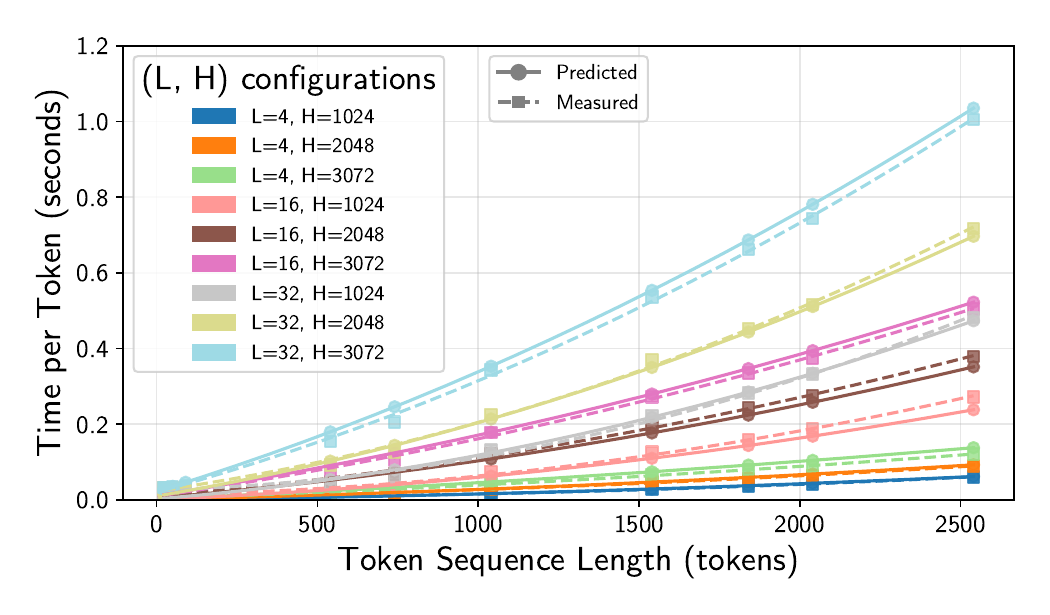}
  \caption{\textbf{Predicted token-generation times (generated by a predictor trained on Llama~1B) vs. ground-truth times (obtained from inference on the Phi-3.5-mini~3.8B model).}}
  \label{fig:phi-pred-res}
\end{figure}

\begin{figure}[t]
  \centering
  \includegraphics[width=\columnwidth]{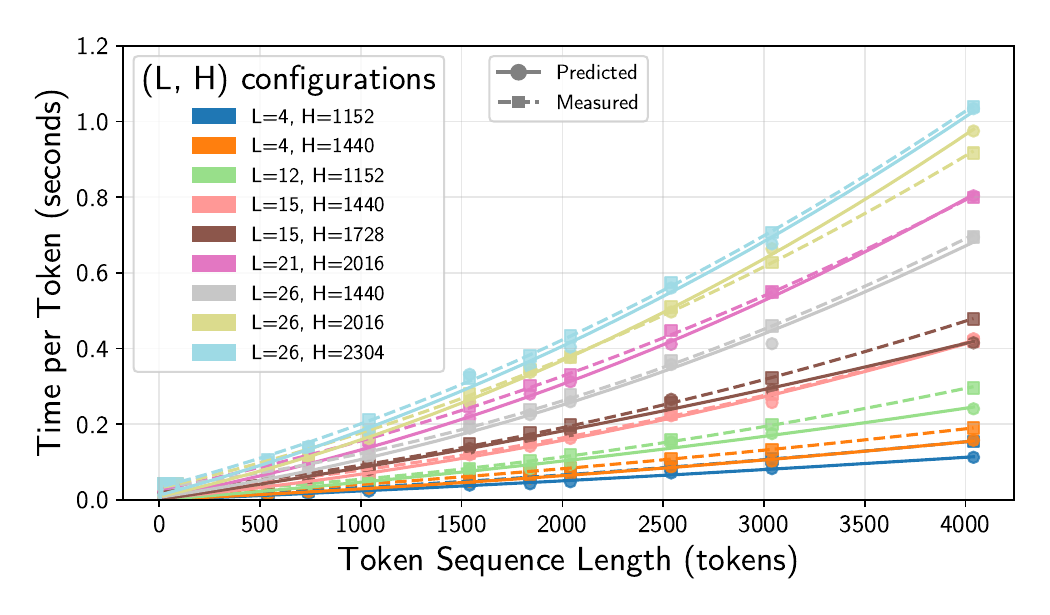}
  \caption{\textbf{Predicted token-generation times (generated by a predictor trained on Llama~1B) vs. ground-truth times (obtained from inference on the Gemma2~2B model).}}
  \label{fig:gemma-pred-res}
\end{figure}

As shown in Figure~\ref{fig:qwen-pred-res}, Figure~\ref{fig:phi-pred-res}, and Figure~\ref{fig:gemma-pred-res}, the predictor accurately captures the timing behavior of different target open models. The quadratic scaling patterns learned by the \texttt{Eager
(Corrected)} predictor on the Llama~1B training dataset generalizes well to per-token generation time prediction across different model families, including Qwen2.5~1.5B, Phi-3.5-mini~3.8B, and Gemma2~2B. This holds despite differences in model architectures and parameter scales. This ability to generalize across model families further supports our argument that the generalizability of linear regression is one of the key reasons for adopting it as the foundation of our prediction approach.

\subsection{Modeling FlashAttention}
\label{sec:appendix-flash}

\niparagraph{\textbf{Runtime Scaling Terms}.}
Below, we derive the terms for the FlashAttention2 kernel based on its original description~\cite{dao2023flashattention2fasterattentionbetter}.

% \noindent
% \textbf{FlashAttention block}:

\begin{equation}
\begin{split}
time_{Flash} &= \alpha_1 T H^2 L d + \alpha_2 T^2 H L d \\
           &\quad + \alpha_3 T^2 D^2 L M^{-1} b c + \alpha_4 T H L b c \\
           &\quad  + \alpha_5 H^2 L b c + \alpha_6 D H L b c
\end{split}
\end{equation}

\begin{equation}
time_{Flash} \sim T^2, D^2, H^2, M^{-1}, L
\end{equation}

In these scaling terms, $D$ denotes the head dimension of the model, and $M$ represents the size of the GPU SRAM. In our experiments on the A10 GPU, we set $M$ to 96~KB. For the experiments on the B200 GPU, we increase $M$ to 228~KB.

\begin{figure}[t]
  \centering
  \includegraphics[width=\columnwidth]{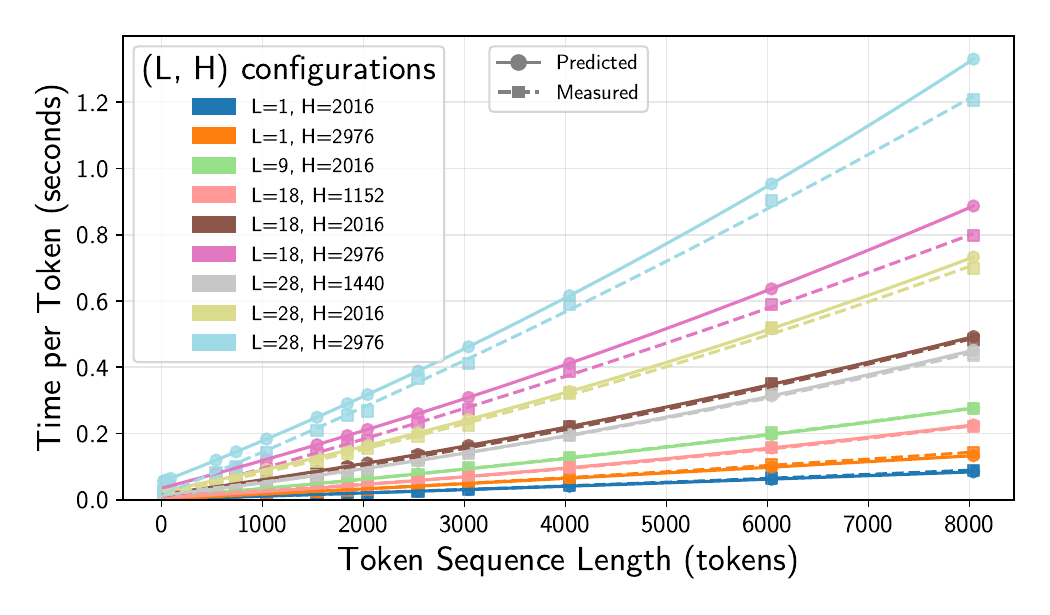}
  \caption{\textbf{Predicted token-generation times vs. ground-truth times for FlashAttention2 kernel.}}
  \label{fig:flash-pred-res}
\end{figure}

\niparagraph{\textbf{Prediction Accuracy}.}
Figure~\ref{fig:flash-pred-res} shows the per-token generation times predicted by our \texttt{Flash2} predictor for different architectural configurations, which closely follow the measured FlashAttention inference times. These results demonstrate that our configurable prediction approach can be extended to support various implementations of the transformer architecture by adjusting a small number of runtime scaling terms, without requiring a new predictor design.

\subsection{Modeling KV-Cache}
\label{sec:appendix-flash-kv}

\niparagraph{\textbf{Runtime Scaling Terms}.}
We define two sets of runtime scaling terms: one for the prefill stage and one for the decoding stage of inference. For the prefill stage, we use the same scaling terms as in \S\ref{sec:appendix-flash}. We then revise the scaling terms for all transformer components to account for decoding-time execution, where each pass processes a single token instead of the full sequence, and attention is computed using cached key-value representations rather than recomputing them. Below, we describe the runtime scaling terms for the decoding stage of KV-cache–enabled inference.

\noindent
\textbf{attention subcomponent}:

\begin{equation}
\begin{split}
time^{decode}_{Att} &= \alpha_1 H^2 L d + \alpha_2 T H L d \\
           &\quad + \alpha_3 T H L b c + \alpha_4 H^2 L b c \\
           &\quad + \alpha_5 H D L b c + \alpha_6 H^2 L b c \\
           &\quad +\alpha_7 H L b c
\end{split}
\end{equation}

\begin{equation}
time^{decode}_{Att} \sim T, H^2, D, L
\end{equation}

\noindent
\textbf{MLP subcomponent}:

\begin{equation}
\begin{split}
time^{decode}_{MLP} &= \alpha_1 HILd + \alpha_2 HILbc \\
           &\quad + \alpha_3 ILbc + \alpha_4 HLbc
\end{split}
\end{equation}

\begin{equation}
time^{decode}_{MLP} \sim H^2, I, L
\end{equation}

\noindent
\textbf{add and normalization subcomponent}:

\begin{equation}
\begin{split}
time^{decode}_{Norm} &= \alpha_1 HLd + \alpha_2 HLbc
\end{split}
\end{equation}

\begin{equation}
time^{decode}_{Norm} \sim H, L
\end{equation}

\noindent
\textbf{embedding/projection (naive) subcomponent}:

\begin{equation}
\begin{split}
time^{decode}_{EmbedProj} &= \alpha_1 HVd + \alpha_2 HVbc
\end{split}
\end{equation}

\begin{equation}
time^{decode}_{EmbedProj} \sim H, V
\end{equation}

\begin{figure}[t]
  \centering

  \begin{subfigure}[t]{0.49\columnwidth}
      \centering
      \includegraphics[width=\textwidth]{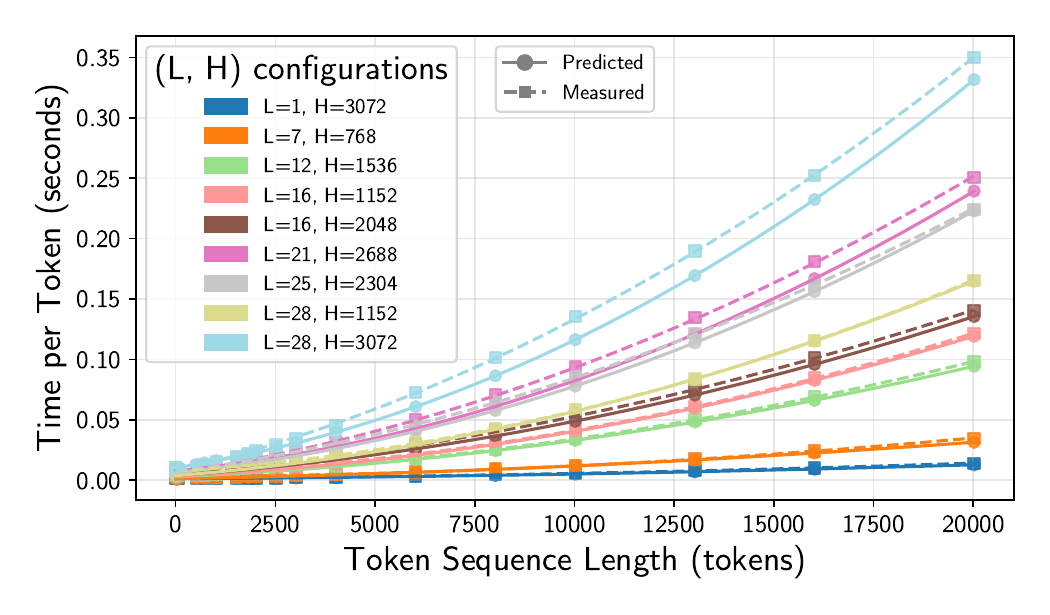}
      \caption{Prefill}
      \label{fig:flash-kv-prefill-res}
  \end{subfigure}
  \hfill
  \begin{subfigure}[t]{0.49\columnwidth}
      \centering
      \includegraphics[width=\textwidth]{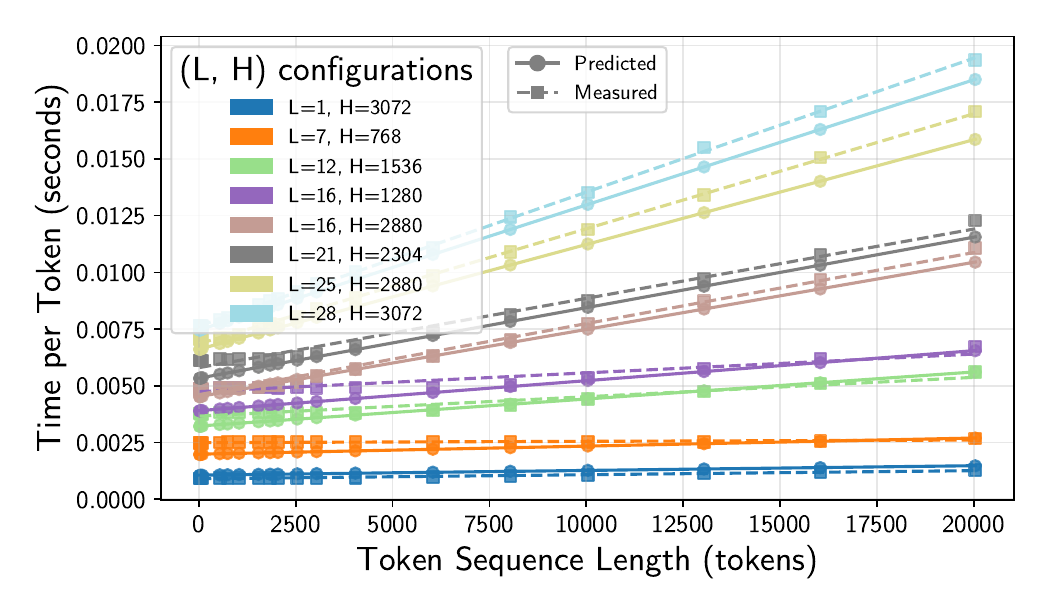}
      \caption{Decode (averaged)}
      \label{fig:flash-kv-decode-res}
  \end{subfigure}
  \caption{\textbf{Predicted token-generation times vs. ground-truth times for KV-cache-enabled inference using FlashAttention2 kernel.}}
  \label{fig:flash-kv-res}
\end{figure}

\niparagraph{\textbf{Prediction Accuracy}.}
Figure~\ref{fig:flash-kv-res} shows the performance of our \texttt{KV-Flash2} predictor across different architectural configurations. Figures~\ref{fig:flash-kv-prefill-res} and~\ref{fig:flash-kv-decode-res} depict per-token generation times for the prefill and decoding stages, respectively. We observe that the hybrid runtime predictor accurately captures both the quadratic dependence on sequence length in the prefill stage and the linear behavior in the decoding stage. These results demonstrate that our runtime prediction framework can be extended to incorporate widely used inference-time optimization techniques, such as KV caching, with minimal modifications.

\subsection{Grid Search Dimensions}
\label{appendix:grid}

We evaluated the following grids, comprising 1540 unique configurations, during the online phase evaluations in \S\ref{sec:attack2-results}, where the only unknown parameters are $H$ and $L$:

\begin{CompactItemize}
    \item Hidden size
    \[
    \begin{split}
    H \in \{32, 64, 128, 256, 512, 640, 768, 896, 960,\\
    1024, 1152, 1248, 1280, 1408, 1440, 1536, 1664,\\
    1792,1824, 1920, 2016, 2048, 2112, 2240, 2304,\\
    2432, 2560, 2688, 2880, 2944, 3072, 3168, 3200,\\
    3264, 3360, 3456, 3582, 3710, 3840, 3968, 4096,\\
    4608, 5120, 6144\}
    \end{split}
    \]
    \item Number of layers %$\quad L \in \{1, 2, \ldots, 35\}$
    \[
    L \in \{1, 2, \ldots, 35\}
    \]
\end{CompactItemize}

When we fix all architectural parameters except one, we expand the search space for the target parameter to reduce bias induced by the discrete values covered in the grid on retrieval accuracy. The grid includes 47 distinct values for $H$ and 44 distinct values for $L$. The resulting ranges for $H$ and $L$ are as follows:

\begin{CompactItemize}
    \item Hidden size
    \[
    \begin{split}
    H \in \{32, 64, 128, 256, 512, 640, 768, 896, 960,\\
    1024, 1152, 1248, 1280, 1408, 1440, 1536, 1664,\\
    1792,1824, 1920, 2016, 2048, 2112, 2240, 2304,\\
    2432, 2560, 2688, 2880, 2944, 3072, 3168, 3200,\\
    3264, 3360, 3456, 3582, 3710, 3840, 3968, 4096,\\
    4352, 4608, 4864, 5120, 5632, 6144\}
    \end{split}
    \]
    \item Number of layers %$\quad L \in \{1, 2, \ldots, 35\}$
    \[
    L \in \{1, 2, \ldots, 45\}
    \]
\end{CompactItemize}

For the remote API experiment, we use the same expanded parameter ranges as in the single-parameter retrieval setting, resulting in a search space of 2,068 $(H, L)$ configurations.

\section{Open Science}
To support the replicability of our results, we commit to making all artifacts related to this research publicly available. The scripts used to run the LLM inference, make the API calls, measure the timings, and analyze the experiment results are all accessible through a public repository (i.e.,
\url{https://anonymous.4open.science/r/LeakyLMs-615B/}), and we provide detailed documentation to facilitate their use by other researchers. We organized the repository to reflect the structure of the paper, allowing readers to easily locate the experimental setup corresponding to each attack and its subsections.

\end{document}